\documentclass[aps,prd,onecolumn,groupedaddress,showpacs,nofootinbib,amssymb]{revtex4}
\usepackage[dvips]{graphicx}
\usepackage{amssymb}
\usepackage{amsmath}
\usepackage{graphicx}
\usepackage{amsfonts}
\usepackage{bm}
\begin{document}

\title{Singular deformations of nearly $R^2$ inflation potentials}
\author{
S.~D.~Odintsov,$^{1,2,4}$\,\thanks{odintsov@ieec.uab.es}
V.~K.~Oikonomou,$^{3,4,5}$\,\thanks{v.k.oikonomou1979@gmail.com}}
\affiliation{
$^{1)}$Institut de Ciencies de lEspai (IEEC-CSIC),
Campus UAB, Carrer de Can Magrans, s/n\\
08193 Cerdanyola del Valles, Barcelona, Spain\\
$^{2)}$ ICREA,
Passeig LluA­s Companys, 23,
08010 Barcelona, Spain\\
$^{3)}$ Department of Theoretical Physics, Aristotle University of Thessaloniki,
54124 Thessaloniki, Greece\\
$^{4)}$ National Research Tomsk State University, 634050 Tomsk, Russia \\
$^{5)}$ Tomsk State Pedagogical University, 634061 Tomsk, Russia\\
}

\begin{abstract}
We investigate in which cases a singular evolution with a singularity of Type IV, can be consistently incorporated in deformations of the $R^2$ inflationary potential. After demonstrating the difficulties that the single scalar field description is confronted with, we use a general two scalar fields model without other matter fluids, to describe the Type IV singular evolution, with one of the two scalar fields being canonical. By appropriately choosing the non-canonical scalar field, we show that the canonical scalar field corresponds to a potential that is nearly the $R^2$ inflation potential. If the Type IV singularity occurs at the end of inflation, the Universe's dynamical evolution near inflation is determined effectively by the canonical scalar field and at late-time the evolution is effectively determined by the non-canonical scalar. We also discuss the evolution of the Universe in terms of the effective equation of state and we show that the Type IV singularity, that occurs at the end of inflation, drives late-time acceleration. If however the singularity occurs at late-time, this might affect the inflationary era. We also investigate which Jordan frame pure $F(R)$ gravity corresponds to the nearly $R^2$ inflation scalar potentials we found. The stability of the solutions in the two scalar fields case is also studied and also we investigate how Type IV singularities can be incorporated in certain limiting cases of $R+R^p$ gravity in the Einstein frame. Finally, we briefly discuss a physical appealing scenario triggered by instabilities in the dynamical system that describes the evolution of the scalar fields.
\end{abstract}

%PACS numbers: 04.50.Kd, 95.36.+x, 98.80.-k, 98.80.Cq

\maketitle

%\makeatletter
%\renewcommand{\theequation}{\Roman{section}\,\Alph{subsection}.\arabic{equation}}
%\@addtoreset{equation}{subsection}{section}
%\makeatother

%\makeatletter
%\renewcommand{\theequation}{\Roman{section}.\arabic{equation}}
%\@addtoreset{equation}{section}
%\makeatother

\def\pp{{\, \mid \hskip -1.5mm =}}
\def\cL{\mathcal{L}}
\def\be{\begin{equation}}
\def\ee{\end{equation}}
\def\bea{\begin{eqnarray}}
\def\eea{\end{eqnarray}}
\def\tr{\mathrm{tr}\, }
\def\nn{\nonumber \\}
\def\e{\mathrm{e}}

\section{Introduction}

One of the most difficult and challenging problems in modern cosmology is the consistent explanation of singularities of any sort. Particularly, we need to explain if these singularities belong to the very own fabric of spacetime, or these are possible indicators that a quantum theory of gravity needs to take place of the classical theories that predict these singularities. Among singularities, there exists a classification that hierarchically determines which are catastrophic, or of crushing type, and which are milder. The most well known singularity of crushing type is the initial singularity, at which the spacetime is geodesically incomplete, meaning that the null and time-like geodesics cannot be continuously extended to arbitrary values of their parameters. These singularities were classified in the pioneer paper of Hawking and Penrose \cite{hawkingpenrose}, were the strong energy conditions theorems were firstly developed. On the other hand, milder singularities are singular spacetime points, at which the strong energy conditions are not violated, although some observables may blow-up at these points. These were first considered in detail in \cite{barrowsing1}, but see also \cite{barrowsing2,barrow,Barrow:2015ora} for relevant studies, and they are known as sudden singularities. In the same line of research, finite-time cosmological singularities \cite{Nojiri:2005sx} belong to the category of non-crushing types of singularities, with the only exception being the Big Rip \cite{Nojiri:2005sx,Caldwell:2003vq,ref5}. At these mild singularities, the fact that some observables blow-up does not necessarily imply geodesics incompleteness. 

One of the most ''harmless'' singularity, is the Type IV singularity, which occurs if the scale factor, the effective energy density and the effective pressure are finite, but the higher derivatives of the Hubble rate diverge. In the recent studies \cite{Nojiri:2015fra,nojirinew}, we extensively studied the effects of a singularity of this type, to the cosmological evolution of a Universe filled with one \cite{Nojiri:2015fra} or two scalar fields \cite{nojirinew}, and particularly on certain inflationary models \cite{inflation}. For a related study on this account, see \cite{Barrow:2015ora}, were similar issues are discussed in the context of sudden singularities. Using some standard reconstruction techniques, applicable to general scalar-tensor theories, we were able to demonstrate that the Type IV singularities can in some cases be consistently incorporated to the cosmological evolution of a Universe filled with one or two scalars and in some other cases, the existence of a Type IV singularity can prove to be catastrophic, affecting explicitly the observables, as these were measured by the Planck \cite{planck} and BICEP2 \cite{BICEP2} collaborations. The very interesting problem is then: can some (well-known) inflationary cosmologies transit to nearby singular inflationary cosmologies, with nearly the same values of inflationary parameters?

The purpose of this paper is to address the issue of having a singular cosmological evolution, for a scalar potential corresponding to the $R^2$ inflation potential, or to a nearly $R^2$ inflation potential in the Einstein frame. In this case, the problem becomes more involved and as we explicitly demonstrate, the need for two scalars becomes compulsory. Having at hand two scalar fields, the possibilities of a successful description of the desired cosmological evolution becomes less involved. For relevant studies on the attributes of the two scalar field cosmological models, the reader is referred to \cite{Nojiri:2005pu,Ito:2011ae}. As we shall evince, by appropriately choosing the scalar fields, one of which is canonical (the inflaton) and one of these non-canonical, the Type IV singular evolution can be incorporated in the context of some variants of $R^2$ inflation potential, with the inflation era being controlled by exactly this nearly $R^2$ inflationary potential, while the late-time era by the non-canonical scalar field. We achieve this by abandoning completely the slow-roll conditions for the second scalar field, and we numerically solve the equations of motion of the non-canonical scalar in order to further support our claims. The model itself has some appealing attributes with regards to it's cosmological implications. Particularly, as we shall show, if we assume that the singularity occurs during the inflationary era, this singularity affects the late-time evolution, with the latter being of nearly phantom type \footnote{Note that it is not hard to construct the model with two scalars where both scalars are of quintessence type so unifying quintessential inflation with Type IV singularity after it, and quintessential dark energy epoch with some of soft finite-time singularities after it. A review on non-singular unified evolution of quintessential inflation with dark energy era is given in \cite{saridakis}.}, but infinitely close to de Sitter. For studies on phantom dark energy era and the possibility of crossing the phantom divide, see \cite{Nojiri:2005pu,ref2,vikman}. In addition, for the possibility of having phantom inflation, see \cite{Liu:2012iba}. In view of the astrophysical data that strongly indicate a phantom dark energy era, this could be of importance and we study this possibility in detail. In addition, we examine which $F(R)$ gravity can generate the corresponding nearly $R^2$ inflationary potentials in the Jordan frame and we also examine the stability of the cosmological solution, in the context of scalar-tensor theories. Moreover, we investigate how the Type IV singularities can be consistently incorporated to limiting cases of $R+R^p$ gravity, in the Einstein frame. Furthermore, we investigate which kinetic term and scalar potential govern the cosmological evolution in the presence of perfect matter fluids. We also provide some strong motivation for the significance of the Type IV singularity and finally we demonstrate that the two scalar field formalism for the observational indices can lead to the same results as in the single scalar field case, thus validating our approximations. 

An important remark is in order. The models we shall use are two scalar field models with one being canonical and the other one non-canonical. The canonical scalar part is chosen to be the $R^2$ inflation model in the Einstein frame. The reason for that is that the model is at $95\%$ concordance with the current observational data. This provides clear motivation for using this model, and it cannot be considered as a toy model. However, the second scalar field is chosen in the simplest way so that a singular evolution is realized. This is a matter of choice, but for simplicity we chose it in the simplest way we could. In this way we shall demonstrate that, although a singular evolution cannot be incorporated in the single scalar field formalism, with two scalar fields this is possible.

This paper is organized as follows: In section II, we briefly provide all the essential information for the finite time singularities and in section III we analyze in detail why within the context of the single scalar field reconstruction method, it is a formidable task to incorporate a Type IV singularity in a $R^2$-like inflation potential. In section IV we address the problem of singular nearly $R^2$ evolution, using two scalar fields. After providing all necessary information, we proceed to the presentation of the model and also we find the detailed form of the nearly $R^2$ potentials in the Einstein frame that can incorporate Type IV singularities. Moreover we find the corresponding Jordan frame $F(R)$ gravity and also we support numerically the assumptions we made during the presentation of the model.  In section V we study the implications of our model, on the cosmological evolution, in terms of the equation of state (EoS), and discuss the appealing possibility that the late time era may be driven by the Type IV singularity which is assumed to occur at the end of the inflationary era. The stability of the cosmological solution is examined in section VI, while in section VII we study some limiting cases of $R+R^p$ gravity in the Einstein frame and the possibility of incorporating a Type IV singularity in the corresponding Einstein frame potentials. In section VIII we study the cosmological evolution with two scalar fields in the presence of perfect matter fluids. In section IX we demonstrate why a Type IV singularity is significant to study, by investigating some implications to the cosmological evolution. Particularly, we briefly analyze qualitatively an interesting possibility that occurs when the dynamical systems that describe the evolution develop instabilities, in the presence of Type IV singularities. In section X, we calculate the observational indices using the two scalar field formalism, and we demonstrate that the approximations we did in the previous sections are valid, for the choice of the parameters we made. The conclusions along with a discussion on the results follow in the end of the paper.

\section{Finite-time singularities classification and conventions}

Before we get into the problem, it is worth recalling here the classification of the finite-time singularities, according to Refs.~\cite{Nojiri:2005sx,sergnoj}.  The finite-time future cosmological singularities are classified as follows,
\begin{itemize}
\item Type I (``Big Rip Singularity'') : When the cosmic time approaches $t \to t_s$, the scale factor $a$,
the effective energy density $\rho_{\mathrm{eff}}$ and also the effective pressure
$p_\mathrm{eff}$ diverge, that is, $a \to \infty$, $\rho_\mathrm{eff} \to \infty$, and
$\left|p_\mathrm{eff}\right| \to \infty$.
For a detailed presentation of the Big Rip singularity, see for example Ref.~\cite{Caldwell:2003vq} and in addition Refs.~\cite{Nojiri:2005sx,ref5}.
\item Type II (``Sudden Singularity'') \cite{barrowsing2,barrow}: When the cosmic time approaches $t \to t_s$, the scale factor $a$ and the effective energy density $\rho_{\mathrm{eff}}$ are finite, that is, $a \to a_s$, $\rho_{\mathrm{eff}} \to \rho_s$. On the contrary, the effective pressure $p_\mathrm{eff}$ diverges, that is, $\left|p_\mathrm{eff}\right| \to \infty$.
\item Type III : When the cosmic time approaches $t \to t_s$, the scale factor is finite, that is, $a \to a_s$, but both the effective pressure and the effective energy density diverge, that is, $\left|p_\mathrm{eff}\right| \to \infty$ and $\rho_\mathrm{eff} \to \infty$.
\item Type IV : This type of singularity is the less harmful, with regards to geodesic incompleteness point of view. For a detailed account on this see \cite{Nojiri:2005sx}. This type of singularity occurs when, as the cosmic time approaches $t \to t_s$, the scale factor, the effective energy density, and
the effective pressure are finite, that is $a \to a_s$, $\rho_\mathrm{eff} \to \rho_s$,
$\left|p_\mathrm{eff}\right| \to p_s$, but some higher derivatives of the Hubble parameter diverge $H\equiv \dot a/a$
\end{itemize}
For a detailed analysis on all the finite-time cosmological singularities, the reader is referred to Ref.~\cite{Nojiri:2005sx}. In all the following considerations, the spacetime metric is assumed to be a spatially flat Friedmann-Robertson-Walker (FRW), of the following form, 
\be
\label{JGRG14}
ds^2 = - dt^2 + a(t)^2 \sum_{i=1,2,3} \left(dx^i\right)^2\, ,
\ee
In terms of the Hubble rate, the effective energy density $\rho_\mathrm{eff} $ and the effective pressure
$p_\mathrm{eff}$ are given as follows,
\be
\label{IV}
\rho_\mathrm{eff} \equiv \frac{3}{\kappa^2} H^2 \, , \quad
p_\mathrm{eff} \equiv - \frac{1}{\kappa^2} \left( 2\dot H + 3 H^2
\right)\, .
\ee

\section{Problem of constructing singular Einstein frame $R^2$ gravity potential with single scalar field}

The purpose of this article is to investigate if it is possible to consistently incorporate a Type IV singularity to potentials that in the Einstein frame are nearly the $R^2$ inflation potential \cite{barrownewrefs,starobinsky}, which, as was indicated by the latest Planck data, can generate viable inflation \cite{planck}. Notice that the scalar-tensor potential version of the $R^2$ inflation were studied for the first time in \cite{barrownewrefs}. Furthermore, the transition between non-singular nearly $R^2$ inflation and singular nearly $R^2$-inflation remains to be the open question. The Jordan frame $R^2$ gravity in the absence of matter fluids, is described by the following four dimensional action, 
\begin{equation}
\label{action1dseenaenase}
\mathcal{S}=\frac{1}{2\kappa^2}\int \mathrm{d}^4x\sqrt{-\hat{g}}\left(R+\frac{R^2}{6M^2}\right)\, ,
\end{equation}
where $\kappa^2=\frac{1}{M_{pl}^2}$, with $M_{pl}=1.22 \times 10^{19}$GeV, and in order to have consistency with recent Planck data, we must set $M\simeq 10^{13}$GeV. By performing the conformal transformation $g_{\mu \nu}=f(\varphi)\hat{g}_{\mu \nu}$, with $f(\varphi)=e^{\sqrt{\frac{2}{3}}\kappa \varphi}$, we may obtain the corresponding Einstein frame scalar-tensor theory, with action,
\be
\label{ma7einframe}
S=\int d^4 x \sqrt{-g}\left\{\frac{1}{2\kappa^2}R - \frac{1}{2}\partial_\mu \varphi
\partial^\mu\varphi - V(\varphi) \right\}\, .
\ee
with $V(\varphi)$, the scalar potential which is,
\begin{equation}\label{astarobisnkypotential}
V(\varphi)=\frac{3}{4}M^2M_{pl}^2\left(1-e^{-\sqrt{\frac{2}{3}\kappa \varphi}}\right)^2
\end{equation}

Note that $R^2$ inflation from the action (\ref{action1dseenaenase}) is equivalent to tree-level non-minimal Higgs inflation \cite{shaposhnikov} in vacuum. Of course, the account of one-loop renormalization group improved corrections in non-minimal Higgs inflation \cite{newrefsserg}, or even two-loop corrections \cite{newrefsserg1}, completely changes the potential and breaks the aforementioned equivalence even in vacuum.

In Refs. \cite{Nojiri:2015fra,nojirinew}, a consistent incorporation of finite-time singularities for power-law potentials was achieved by using one or two scalar fields. In the present work we shall attempt to do the same for the potential (\ref{astarobisnkypotential}). In the approach we adopted in Ref. \cite{Nojiri:2015fra,nojirinew}, we made use of general scalar-tensor theories with one or two scalars, but for the potential (\ref{astarobisnkypotential}), it is not easy to incorporate a Type IV singularity by using only one scalar field, as we now explicitly demonstrate. Therefore, the need for two scalar fields is compelling. In addition, the use of two scalar fields has very appealing cosmological consequences, since a unified description of singular Type IV inflation described by the potential (\ref{astarobisnkypotential}), and of a nearly phantom dark energy era, is possible in the theoretical framework we shall use. More importantly, it is possible that the dark energy era is driven by the finite-time singularity it self, which is assumed to occur at the end of inflation.  

Before we proceed to the description of the study with two scalar fields, let us see why a single scalar-tensor theory fails to consistently incorporate the Type IV singularity. Consider the scalar field action given below,
\be
\label{ma7}
S=\int d^4 x \sqrt{-g}\left\{
\frac{1}{2\kappa^2}R - \frac{1}{2}\omega(\phi)\partial_\mu \phi
\partial^\mu\phi - V(\phi)  \right\}\, .
\ee
which describes a single, non-canonical scalar field. The function $\omega(\phi)$ is the kinetic function and $V(\phi)$ is the corresponding scalar potential and also we assume a flat FRW background of the form (\ref{JGRG14}). Therefore, the energy density and the pressure are equal to,
\be
\label{ma8}
\rho = \frac{1}{2}\omega(\phi){\dot \phi}^2 + V(\phi)\, ,\quad
p = \frac{1}{2}\omega(\phi){\dot \phi}^2 - V(\phi)\, .
\ee
and consequently, the scalar potential $V(\phi)$ and the kinetic term $\omega(\phi)$ can be written in terms of the Hubble parameter as follows,
\be
\label{ma9}
\omega(\phi) {\dot \phi}^2 = - \frac{2}{\kappa^2}\dot H\, ,\quad
V(\phi)=\frac{1}{\kappa^2}\left(3H^2 + \dot H\right)\, .
\ee
From action (\ref{ma7}), after making the following transformation,
\be
\label{ma13}
\varphi \equiv \int^\phi d\phi \sqrt{\omega(\phi)} \,,
\ee
it is possible to rewrite the action (\ref{ma7}) in terms of a canonical scalar field $\varphi$. Indeed, the kinetic term of the scalar field becomes, 
\be
\label{ma13b}
 - \omega(\phi) \partial_\mu \phi \partial^\mu\phi
= - \partial_\mu \varphi \partial^\mu\varphi\, .
\ee
Therefore, the action (\ref{ma7}) becomes, 
\be
\label{ma7einfram12e}
S=\int d^4 x \sqrt{-g}\left\{
\frac{1}{2\kappa^2}R - \frac{1}{2}\partial_\mu \varphi
\partial^\mu\varphi - V(\varphi) \right\}\, .
\ee
In order to study the incorporation of finite-time singularities, in Refs. \cite{Nojiri:2015fra,nojirinew} the scalar-reconstruction method was used \cite{Nojiri:2005pu,Capozziello:2005tf}, which we now describe in brief. In the context of the scalar-reconstruction method, it is assumed that both the kinetic term $\omega(\phi)$ and the scalar potential $V(\phi)$, are written in terms of a function $f(\phi)$, as follows,
\be
\label{ma10}
\omega(\phi)=- \frac{2}{\kappa^2}f'(\phi)\, ,\quad
V(\phi)=\frac{1}{\kappa^2}\left(3f(\phi)^2 + f'(\phi)\right)\, ,
\ee
Consequently, neglecting the contribution of matter fluids, the FRW equations (\ref{ma8}), can be written as,
\be
\label{ma11}
\phi=t\, ,\quad H=f(t)\, .
\ee
The difficulty in incorporating finite-time singularities for the potential (\ref{astarobisnkypotential}) is traced on the exponential form of the potential. In order a finite-time singularity occurs, the non-canonical scalar-tensor reconstruction function $f(t)$ must certainly be of the form,
\begin{equation}\label{f}
f(t)=c(t-t_s)^b+f_1(t)
\end{equation}
where the function $f_1(t)$ is the corresponding non-canonical scalar-tensor reconstruction function that produces the scalar potential (\ref{astarobisnkypotential}), when canonically transformed of course. In addition to (\ref{f}), there exist similar forms for the reconstruction function $f(t)$, for example,
\begin{equation}\label{newf}
f(t)=f_1(t)(t-t_s)^b ,{\,}{\,}{\,}f(t)=e^{(t-t_s)^{\alpha}}+c
\end{equation}
with $c$ an arbitrary constant. In all the above cases, it is really difficult to incorporate these in the potential (\ref{astarobisnkypotential}), for the following two reasons:

\begin{itemize}

\item At first, the function $f(t)$ must be such, so that the integral $\varphi =\int^{\phi} \sqrt{f'(\phi )} \mathrm{d}\phi$, can be solved explicitly in terms of $\phi =\phi (\varphi)$, which is a rather formidable task for the functional forms (\ref{f}) and (\ref{newf}).

\item Secondly, if someone discovers this $f(\phi)$ function, so that $\phi =\phi (\varphi)$ is explicitly solved, then the scalar-tensor potential,  when expressed in terms of the canonical scalar field, becomes very constrained. Indeed, suppose $\phi(\varphi)$ is found explicitly, then the potential $V(\phi (\varphi))$ is given by
\be
\label{ma100101}
V(\phi (\varphi))=\frac{1}{\kappa^2}\left(3f(\phi(\varphi))^2 + f'(\phi (\varphi))\right)\, ,
\ee
and this must be of the following form,
\begin{equation}\label{defoeme1}
V(\phi (\varphi))=\frac{3}{4}M^2M_{pl}^2\left(1-e^{-\sqrt{\frac{2}{3}\kappa \varphi}}\right)^2+V_1(\phi (\varphi))
\end{equation}
which is very difficult to achieve. 

\end{itemize}
Therefore, the need for an alternative approach is compelling, in order to incorporate Type IV or other singularities in the cosmological evolution. In the next section we shall make use of two scalar fields in order to achieve this.

\section{Singular nearly $R^2$ gravity evolution with two scalar fields}

The use of the two scalar field reconstruction scheme offers many more possibilities of cosmological evolution in comparison to the single scalar field method. In addition, possible inconsistencies that may occur in the single scalar field reconstruction, like for example the appearance of infinite instabilities at the cosmic time corresponding to the phantom- non-phantom transition, are properly amended \cite{Nojiri:2005pu,Capozziello:2005tf,vikman}. For a recent study on this account see also \cite{nojirinew}. We shall adopt the two scalar fields reconstruction scheme and as we shall demonstrate, apart from the consistent incorporation of the Type IV singularity in the theoretical framework of the nearly $R^2$ inflation gravity, interesting cosmological phenomenology is generated. Before getting into the details, we shall describe in brief the theoretical apparatus that we shall make extensive use of.

The two scalar field scalar-tensor action we shall consider has the following form, 
\be
\label{A1}
S=\int d^4 x \sqrt{-g}\left\{\frac{1}{2\kappa^2}R
 - \frac{1}{2}\omega(\phi)\partial_\mu \phi \partial^\mu \phi
 - \frac{1}{2}\eta(\chi)\partial_\mu \chi\partial^\mu \chi
 - V(\phi,\chi)\right\}\, .
\ee
In the above equation, the function $\omega(\phi)$ represents the kinetic function corresponding to the scalar field $\phi$, and  $\eta(\chi)$ is the kinetic function corresponding to the other scalar field $\chi$. It is to be understood that in the case one of the kinetic functions $\omega(\phi)$ or $\eta(\chi)$ is negative, then the corresponding scalar field becomes a ghost (phantom) field. For simplicity, we assume that the non-canonical scalar fields $\phi$ and $\chi$ depend only on the cosmic time $t$. For a spatially flat FRW metric of the form (\ref{JGRG14}), the FRW equation corresponding to the action (\ref{A1}) are, 
\be
\label{A2}
\omega(\phi) {\dot \phi}^2 + \eta(\chi) {\dot \chi}^2
= - \frac{2}{\kappa^2}\dot H\, ,\quad
V(\phi,\chi)=\frac{1}{\kappa^2}\left(3H^2 + \dot H\right)\, .
\ee
If the generalized scalar potential $V(\phi,\chi)$ and the kinetic functions $\omega(\phi)$, $\eta(\chi)$ satisfy the following, 
\be
\label{A3}
\omega(t) + \eta(t)=- \frac{2}{\kappa^2}f'(t)\, ,\quad
V(t,t)=\frac{1}{\kappa^2}\left(3f(t)^2 + f'(t)\right)\, ,
\ee
then, the explicit solution of Eqs.~(\ref{A2}) has the following form, 
\be
\label{A4}
\phi=\chi=t\, ,\quad H=f(t)\, .
\ee
This method materializes in brief the two scalars reconstruction scheme, a detailed account of which can be found in \cite{Nojiri:2005pu}.  An attribute of this method is that in principle there is much more freedom in the choice of the kinetic functions $\omega (\phi)$ and $\eta (\chi)$. A very convenient choice for these functions is, 
\be
\label{A5}
\omega(\phi) =-\frac{2}{\kappa^2}\left\{f'(\phi)
- \sqrt{\alpha(\phi)^2 + f'(\phi)^2} \right\}\, ,\quad
\eta(\chi) = -\frac{2}{\kappa^2}\sqrt{\alpha(\chi)^2 + f'(\chi)^2}\, . 
\ee
with the function $\alpha (x)$ being an arbitrary function of the scalar fields. We define a new auxiliary function $\tilde f(\phi,\chi)$ to be of the following form, 
\be
\label{A6}
\tilde f(\phi,\chi)\equiv - \frac{\kappa^2}{2}\left(\int d\phi 
\omega(\phi) + \int d\chi \eta(\chi)\right)\, .
\ee
This function has the important property,
\be
\label{A7}
\tilde f(t,t)=f(t)\, .
\ee
which actually fixes the arbitrary constants of integration arising in Eq.~(\ref{A6}). The scalar potential can be written as a function of $\tilde f(\phi,\chi)$, in the following way, 
\be
\label{A8}
V(\phi,\chi)=\frac{1}{\kappa^2}\left(3{\tilde f(\phi,\chi)}^2
+ \frac{\partial \tilde f(\phi,\chi)}{\partial \phi}
+ \frac{\partial \tilde f(\phi,\chi)}{\partial \chi} \right)\, ,
\ee
Consequently, along with the FRW equations Eq.~(\ref{A2}), the following two field equations hold true,
\be
\label{A9}
0 = \omega(\phi)\ddot\phi + \frac{1}{2}\omega'(\phi) {\dot \phi}^2
+ 3H\omega(\phi)\dot\phi + \frac{\partial \tilde V(\phi,\chi)}{\partial 
\phi} \, , \quad
0 = \eta(\chi)\ddot\chi + \frac{1}{2}\eta'(\chi) {\dot \chi}^2
+ 3H\eta(\chi)\dot\chi + \frac{\partial \tilde V(\phi,\chi)}{\partial 
\chi}\, .
\ee
The kinetic functions $\omega (\phi)$, $\eta (\chi)$ and the two-scalar potential $V(\phi,\chi)$, define a two-scalar field scalar-tensor theory, the cosmological evolution of which is given by Eq.~(\ref{A4}). In the following we shall focus to theories with a Hubble rate of the form,
\be
\label{IV1Bnoj1}
H(t) = f_1(t) + f_2(t) \left( t_s - t \right)^\alpha\, ,
\ee
with $\alpha$ an arbitrary parameter that actually critically determines the type of the finite-time singularity. Particularly, the classification of the types of finite-time singularities, for various values of the parameter $\alpha$, is given in the list below.
\begin{itemize}\label{lista}
\item $\alpha<-1$ corresponds to the Type I singularity.
\item $-1<\alpha<0$ corresponds to Type III singularity.
\item $0<\alpha<1$ corresponds to Type II singularity.
\item $\alpha>1$ corresponds to Type IV singularity.
\end{itemize}
For the needs of this paper, special emphasis will be given on the Type IV singularity, which is the most harmless, from a geodesics incompleteness point of view. A thorough analysis for the impact of this singularity on single scalar field inflation was performed in \cite{Nojiri:2015fra}. As we already noted, the function $\alpha (x)$ appearing in equation (\ref{A5}), can freely be chosen. We shall assume that in our case it has the following form,
\begin{equation}\label{alphaxnewdef1}
\alpha (x)=\sqrt{\left ( f_2' (x)\left (t_s-x \right )^{\alpha}+\alpha f_2(x)\left (t_s-x\right )^{\alpha-1}\right )^2-\left( H'(x)\right)^2}\, .
\end{equation}
where it is to be understood that the variable $x$ can be either $\phi$ or $\chi$. With this choice of $\alpha(x)$, the resulting expressions for the scalar field kinetic functions $\omega (\phi)$ and $\eta (\chi)$ are very much simplified, and these are of the following form,
\be
\label{B1noj}
\omega(\phi) =-\frac{2}{\kappa^2}f_1'(\phi)\, ,\quad
\eta(\chi) = -\frac{2}{\kappa^2} \left( f_2'(\chi) \left( t_s - \chi \right)^\alpha
+ \alpha f_2(\chi) \left( t_s - \chi \right)^{\alpha - 1} \right)\, , 
\ee
The specific form of the function $\alpha (x)$ appearing in Eq.~(\ref{alphaxnewdef1}), specifies the final form of the auxiliary function $\tilde{f}(\phi,\chi)$, which we defined in Eq.~(\ref{A6}). Particularly, this becomes,
\be
\label{B2noj}
\tilde f(\phi,\chi) = f_1(\phi) + f_2(\chi) \left( t_s - \chi \right)^\alpha\, ,
\ee
Thereby, the two scalar field potential has the final form,
\be
\label{B3noj}
V(\phi,\chi)=\frac{1}{\kappa^2}\left(3\left(
f_1(\phi) + f_2(\chi) \left( t_s - \chi \right)^\alpha \right)^2 
+ f_1'(\phi) +  f_2'(\chi) \left( t_s - \chi \right)^\alpha
+ \alpha f_2(\chi) \left( t_s - \chi \right)^{\alpha - 1} \right)\, .
\ee
Having relations (\ref{B1noj}) and (\ref{B3noj}), we may easily proceed to reconstruct the deformed singular version of nearly $R^2$ inflation. Our aim is to have a Type IV singularity in the Hubble rate, while at the same time the scalar potential contains some deformation of the potential (\ref{astarobisnkypotential}), at least in one of the two scalar fields. In order to achieve this, we assume that the Hubble rate has the following form,
\be
\label{IV1Bnoj}
H(t) =  \frac{c_1}{c_2+c_3 t}+c_4+ c_5 \left( t_s - t \right)^\alpha\, ,
\ee
where the parameters $c_i$, with $i=1,..5$, are arbitrary constant parameters. This is the simplest choice we can make in order to consistently accommodate the Type IV singularity in the nearly $R^2$ inflation cosmological model.
From the classification given in the list above, when $\alpha >1$, the cosmological evolution develops a Type IV singularity and this is what we assume in the rest of this paper. In addition, for reasons that will become clear later, we assume that $\alpha$ has the following form,
\be
\label{IV2}
\alpha= \frac{n}{2m + 1}\, ,
\ee
where $m$ can be any positive integer and $n$ is some positive even integer, chosen in such a way so that $\alpha>1$. For the choice (\ref{IV1Bnoj}), the kinetic functions $\omega (\phi)$ and $\eta (\chi)$ given in Eq. (\ref{B1noj}) become, 
\begin{equation}\label{omeganforhilltopdouble}
\omega (\phi )=\frac{2 c_1 c_3}{\kappa ^2 (c_2+c_3 \phi )^2}\, , \quad 
\eta (\chi)= -\frac{2 c_2 \alpha  (t_s-\chi )^{-1+\alpha }}{\kappa ^2}\, .
\end{equation}
from which it is obvious that the scalar field $\chi$ is a non-phantom scalar. The corresponding $\alpha (x)$ function of Eq. (\ref{alphaxnewdef1}) now becomes,
\begin{equation}\label{alphaxnewdef}
\alpha (x)=\sqrt{c_5^2 (t_s-x)^{-2+2 \alpha } \alpha ^2-\left(-\frac{c_1 c_3}{(c_2+c_3 x)^2}-c_5 (t_s-x)^{-1+\alpha } \alpha \right)^2}\, .
\end{equation}
With the choice (\ref{alphaxnewdef}), the function $\tilde f(\phi,\chi)$ becomes equal to,
\begin{equation}\label{barf}
\tilde f(\phi,\chi)=-c_4+\frac{c_1}{c_2+c_3 \phi }+c_5 (t_s-\chi )^{\alpha }\, ,
\end{equation}
and consequently, the two scalar field potential $V(\phi,\chi)$ becomes equal to,
\begin{align}\label{scalarpotentialhilltop}
& V(\phi,\chi)=\frac{3 c_4^2}{\kappa ^2}+\frac{3 c_1^2}{\kappa ^2 (c_2+c_3 \phi )^2}-\frac{c_1 c_3}{\kappa ^2 (c_2+c_3 \phi )^2}+\frac{6 c_1 c_4}{\kappa ^2 (c_2+c_3 \phi )}
\\ \notag &-\frac{c_5 \alpha  (t_s-\chi )^{-1+\alpha }}{\kappa ^2}+\frac{6 c_4 c_5 (t_s-\chi )^{\alpha }}{\kappa ^2}+\frac{6 c_1 c_5 (t_s-\chi )^{\alpha }}{\kappa ^2 (c_2+c_3 \phi )}+\frac{3 c_5^2 (t_s-\chi )^{2 \alpha }}{\kappa ^2}\, .
\end{align}
Now we perform the transformation (\ref{ma13}) to the scalar field $\phi$, in order to transforms it, to it's canonical scalar field counterpart $\varphi $. By using (\ref{ma13}) and the form for $\omega (\phi )$ given in Eq. (\ref{omeganforhilltopdouble}), we easily obtain the expression that relates the canonical scalar field $\varphi$ with the non-canonical scalar field $\phi$,
\begin{equation}\label{noncanonicalcanonicalrel}
c_2+c_3 \phi = e^{\frac{\kappa }{\sqrt{2c_1}}\varphi }
\end{equation}
Therefore, the two scalar field action (\ref{A1}) becomes,
\be
\label{A1new}
S=\int d^4 x \sqrt{-g}\left\{\frac{1}{2\kappa^2}R
 - \frac{1}{2}\partial_\mu \varphi \partial^\mu \varphi
 -\frac{1}{2}\left(\frac{2 c_2 \alpha  (\chi-t_s )^{-1+\alpha }}{\kappa ^2}\right)\partial_\mu \chi\partial^\mu \chi
 - \tilde{V}(\varphi,\chi)\right\}\, .
\ee
where the potential $\tilde{V}(\varphi,\chi)$ that contains the canonical scalar field $\varphi$, is equal to,
\begin{align}\label{finalpotential}
& \tilde{V}(\varphi,\chi)= \frac{3 c_4^2}{\kappa ^2}+\frac{3 c_1^2 e^{-\frac{2 \kappa  \varphi }{\sqrt{2 c_1}}}}{\kappa ^2 }-\frac{c_1 c_3 e^{-\frac{2 \kappa  \varphi }{\sqrt{2 c_1}}}}{\kappa ^2 }+\frac{6 c_1 c_4 e^{-\frac{ \kappa  \varphi }{\sqrt{2 c_1}}}}{\kappa ^2 }
\\ \notag & -\frac{c_5 \alpha  (t_s-\chi )^{-1+\alpha }}{\kappa ^2}+\frac{6 c_4 c_5 (t_s-\chi )^{\alpha }}{\kappa ^2}+\frac{6 c_1 c_5 (t_s-\chi )^{\alpha } e^{-\frac{ \kappa  \varphi }{\sqrt{2 c_1}}}}{\kappa ^2 }+\frac{3 c_5^2 (t_s-\chi )^{2 \alpha }}{\kappa ^2}
\end{align}
and also we made use of the fact that if $\alpha=n/(2m+1)$, with $n=$even, then $\alpha-1=(n-2m-1)/(2m+1)$, and so $n-2m-1=$odd. Consequently, the following relation holds true in Eq. (\ref{A1new}),
\begin{equation}\label{neweqn}
(t_s-\chi )^{-1+\alpha }=-(\chi-t_s )^{-1+\alpha }
\end{equation} 
The final expression of the two scalar field potential $\tilde{V}(\varphi,\chi)$, when the canonical scalar field $\varphi $ is taken into account, has similar form to some variants of the potential appearing in Eq. (\ref{astarobisnkypotential}). As we now explicitly demonstrate, the corresponding Jordan frame $F(R)$ theories, are some modified versions of $R^2$, the form of which is determined solely by the values of the free parameters $c_i$, with $i=1,..4$.

\subsection{Jordan frame $F(R)$ gravity}

\subsubsection{Model I: Jordan frame $a R^2+R+\Lambda$ gravity}

For notational simplicity we introduce the following three constant parameters $C_0$, $C_1$ and $C_2$, which in terms of the parameters $c_i$, $i=1,..4$ which appear in Eq. (\ref{finalpotential}) are defined to be,
\begin{equation}\label{newparam}
C_0=\frac{3 c_4^2}{\kappa ^2},{\,}{\,}{\,}C_1=\frac{3 c_1^2}{\kappa ^2 }-\frac{c_1 c_3}{\kappa ^2},{\,}{\,}{\,}C_2=\frac{6 c_1 c_4}{\kappa ^2}
\end{equation}
Using this notation, the potential (\ref{finalpotential}) can be written in the following form,
\begin{equation}\label{newformpot}
 \tilde{V}(\varphi,\chi)=V_s(\varphi)-\frac{c_5 \alpha  (t_s-\chi )^{-1+\alpha }}{\kappa ^2}+\frac{6 c_4 c_5 (t_s-\chi )^{\alpha }}{\kappa ^2}+\frac{6 c_1 c_5 (t_s-\chi )^{\alpha } e^{-\frac{ \kappa  \varphi }{\sqrt{2 c_1}}}}{\kappa ^2 }+\frac{3 c_5^2 (t_s-\chi )^{2 \alpha }}{\kappa ^2}
\end{equation}
where we have set $V_s(\varphi )$ to be equal to,
\begin{equation}\label{vsnew1}
V_s(\varphi)= C_0+C_2e^{-\frac{ 2\kappa  \varphi }{\sqrt{2 c_1}}}+C_1e^{-\frac{ \kappa  \varphi }{\sqrt{2 c_1}}}
\end{equation}
In order the potential (\ref{vsnew}) resembles the one appearing in Eq. (\ref{astarobisnkypotential}), we must set $c_1=\frac{3}{4}$. Then, by doing so the potential reads,
\begin{equation}\label{vsnew}
V_s(\varphi)= C_0+C_2e^{-2\sqrt{\frac{2}{3}}\kappa   \varphi }+C_1e^{-\sqrt{\frac{2}{3}}\kappa   \varphi }
\end{equation}
As we explicitly demonstrate, the potential in Eq. (\ref{vsnew}) can be a variant form of the potential (\ref{astarobisnkypotential}). Before doing so, let us set the theoretical framework of our analysis and describe the dynamics of the scalar field $\chi $. The initial conditions of this scalar field can be chosen in such a way so that it starts from significantly small values of $\chi $, which during inflation can be negligible, in comparison to the contribution coming from the scalar field $\varphi $. In order to further suppress the contribution of the scalar field $\chi$ during inflation, we choose $c_5\ll 1$ and also $c_5\ll |c_i|$, with $i=1,..4$. Since $\alpha >1$, as the cosmic time increases, the scalar field $\chi$ grows larger and we can in principle choose the parameter $c_5$ in such a way so that the contribution of the scalar field $\chi$ becomes significant at cosmic times much more later than the ending of inflation. Moreover, we assume that the slow-roll condition does not apply for the scalar $\chi $. In addition to these, we choose $t_s$ in Eq. (\ref{IV1Bnoj}), which is the cosmic time where the singularity occurs, to be exactly the cosmic time when inflation ends. In a later section, we shall thoroughly discuss these choices and possible alternative choices. In order to support our claim that the scalar field evolution goes as we just described, in the end of this section, we shall perform a numerical analysis to see explicitly how the dynamical evolution of the scalar field $\chi$ goes. 

As a consequence of these constraints, during and at the end of the inflationary era, the potential (\ref{newformpot}) is approximately equal to,
\begin{equation}\label{vapprox}
\tilde{V}(\varphi,\chi)\simeq C_0+C_2e^{-2\sqrt{\frac{2}{3}}\kappa   \varphi }+C_1e^{-\sqrt{\frac{2}{3}}\kappa   \varphi }
\end{equation}  
In addition to this, the kinetic term of the scalar $\chi$, during and at the end of the inflationary era, can be disregarded (for reasons we explained above), so that the action (\ref{A1new}) for $t\lesssim t_s$,
\be
\label{A1newnearinfl}
S\simeq \int d^4 x \sqrt{-g}\left\{\frac{1}{2\kappa^2}R
 - \frac{1}{2}\partial_\mu \varphi \partial^\mu \varphi
 - V_s(\varphi)\right\}\, .
\ee
with $V_s(\varphi)$ being defined in Eq. (\ref{vsnew}). So the inflationary dynamics is completely determined by the canonical scalar field $\varphi$. The action (\ref{A1newnearinfl}), is the Einstein frame counterpart action of the Jordan frame pure $F(R)$ gravity \cite{reviews1,reviews2},
\begin{equation}
\label{pure}
\mathcal{S}=\frac{1}{2\kappa^2}\int \mathrm{d}^4x\sqrt{-\hat{g}}\left(-\frac{C_1}{2C_0}R+\frac{R^2}{4C_0}+\Lambda\right)\, ,
\end{equation}
with $\Lambda =\frac{C_1^2}{4C_0}-C_2$. In order to see this, we must conformally transform action (\ref{pure}). The technique is quite well known and for a detailed analysis on this, the reader is referred to \cite{reviews1,reviews2} and also \cite{sergeistarobinsky}. In addition, for a useful study with similar scalar potentials, see Ref. \cite{newref}. Starting from action (\ref{pure}), we introduce the auxiliary field $A$, and the Jordan frame action that describes a pure $F(R)$ gravity, namely,
\begin{equation}\label{purefrgrav}
S=\int \mathrm{d}x^4\sqrt{-\hat{g}}F(R)
\end{equation}
can be written in the following way,
\begin{equation}\label{action1dse111}
\mathcal{S}=\frac{1}{2\kappa^2}\int \mathrm{d}^4x\sqrt{-\hat{g}}\left ( F'(A)(R-A)+F(A) \right ),
\end{equation}
where the $F(R)$ function is,
\begin{equation}\label{frfunction}
F(R)=-\frac{C_1}{2C_0}R+\frac{R^2}{4C_0}+\Lambda
\end{equation}
It can easily be verified that by varying (\ref{action1dse111}) with respect to $A$, yields the solution $A=R$, a fact that validates the mathematical equivalence of the actions (\ref{purefrgrav}) and (\ref{action1dse111}). In order to find the Einstein frame scalar theory, we perform the following canonical transformation,
\begin{equation}\label{can}
\varphi =-\sqrt{\frac{3}{2k^2}}\ln (F'(A))
\end{equation}
where $\varphi$ is the Einstein frame scalaron (or inflaton field). By making the conformal transformation of the Jordan frame metric,
\begin{equation}\label{conftransmetr}
g_{\mu \nu}=e^{-\varphi }\hat{g}_{\mu \nu }
\end{equation}
where the ''hat'' denotes the Jordan frame metric, we easily obtain the following Einstein frame scalar field action,
\begin{align}\label{einsteinframeaction}
& \mathcal{\tilde{S}}=\int \mathrm{d}^4x\sqrt{-g}\left ( \frac{R}{2k^2}-\frac{1}{2}\left (\frac{F''(A)}{F'(A)}\right )^2g^{\mu \nu }\partial_{\mu }A\partial_{\nu }A -\frac{1}{2k^2}\left ( \frac{A}{F'(A)}-\frac{F(A)}{F'(A)^2}\right ) \right ) \\ \notag &
= \int \mathrm{d}^4x\sqrt{-g}\left ( \frac{R}{2k^2}-\frac{1}{2}g^{\mu \nu }\partial_{\mu }\varphi\partial_{\nu }\varphi -V(\varphi )\right )
\end{align}
The potential $V(\varphi )$ as a function of the canonical scalar field $\varphi $ is equal to,
\begin{align}\label{potentialvsigma}
V(\varphi )=\frac{A}{F'(A)}-\frac{F(A)}{F'(A)^2}=\frac{1}{2k^2}\left ( e^{\sqrt{2k^2/3}\varphi }R\left (e^{-\sqrt{2k^2/3}\varphi} \right )- e^{2\sqrt{2k^2/3}\varphi }F\left [ R\left (e^{-\sqrt{2k^2/3}\varphi} \right ) \right ]\right )
\end{align}
The function $R\left (e^{-\sqrt{2k^2/3}\varphi} \right )$ is the solution of Eq. (\ref{can}), with respect to $A$, bearing in mind that $A=R$. Then, if the potential $V(\varphi )$ is known, the corresponding $F(R)$ gravity may easily be obtained from Eq. (\ref{potentialvsigma}) and Eq. (\ref{can}). Indeed by dividing Eq. (\ref{potentialvsigma}) with $e^{2\sqrt{2/3}\kappa \varphi}$ and taking the derivative with respect to the scalar curvature $R$, we obtain the following relation,
\begin{equation}\label{solvequation}
RF_R=-2\kappa^2\sqrt{\frac{3}{2\kappa^2}}\frac{\mathrm{d}}{\mathrm{d}\varphi}\left(\frac{V(\varphi)}{e^{2\left(\sqrt{2\kappa^2/3}\right)\varphi}}\right)
\end{equation}
where $F_R=\frac{\mathrm{d}F(R)}{\mathrm{d}R}$. Combining Eqs. (\ref{solvequation}) and (\ref{can}), for the potential $V_s(\varphi )$ of Eq. (\ref{vsnew}), we obtain the following algebraic equation,
\begin{equation}\label{newalgreaequa}
2C_0F_R^2+C_1F_R-RF_R=0
\end{equation} 
Assuming that $F_R\neq 0$, we easily obtain the solution,
\begin{equation}\label{frsol}
F(R)=-\frac{C_1}{2C_0}R+\frac{R^2}{4C_0}+\Lambda
\end{equation}
which is nearly an $R+aR^2$ Jordan frame gravity. But the constant parameters $C_i$, $i=1,2,3$ must be further specified for consistency, and as a consequence, the parameters $c_i$ $i=1,..4$ are further constrained. In order for the $F(R)$ gravity of Eq. (\ref{frsol}) to be Einstein gravity plus corrections, the coefficient of $R$ must be set equal to one, that is,
\begin{equation}\label{constr1}
\frac{C_1}{2C_0}=-1,{\,}{\,}{\,}
\end{equation}
As a consequence of this, in conjunction with Eq. (\ref{newparam}), these two relations result to $c_1=-c_4$. This requirement indicates that either $c_1$ or $c_4$ must be negative. But since previously we have set $c_1=3/4$, this means that $c_4=-3/4$. Then, in order the Hubble parameter is always positive, the parameters $c_2$ and $c_3$ must be very small, that is $c_2\ll 1$ and $c_3\ll 1$. Therefore for all times, by appropriately choosing the parameters $c_2$ and $c_3$ to be very small, the Hubble rate is always positive. Before we proceed to the next model, we summarize below all the constraints that the parameters $c_i$, $i=1,..5$ must satisfy, at least for this model: 
\begin{equation}\label{constrmodel1}
c_1=-c_4=\frac{3}{4},{\,}{\,}{\,}c_2\ll 1,{\,}{\,}{\,}c_3\ll 1,{\,}{\,}{\,}c_5\ll 1
\end{equation} 
With regards to the later constraint, namely $c_5\ll 1$, specifically, $c_5$ must satisfy,
\begin{equation}\label{ndhdgf}
c_5\ll \frac{3}{4(c_2+c_3t)}
\end{equation}
at all cosmic time values. This constraint will be of some importance in a later section. 

The situation we just studied, referred to the case that $t\simeq t_s$, so it corresponds to the inflationary era. It is to be understood that as the cosmic time increases, the scalar field $\chi$, will at some point dominate during the cosmological evolution, while the potential of the canonical scalar field will be negligible. This will occur because the slow-roll approximation for the exponential potential (\ref{vsnew}), is violated when $\varphi \rightarrow \infty$. Therefore, the action (\ref{A1new}) for $t\gg t_s$ will take the form,
\be
\label{A1newnew}
S\simeq \int d^4 x \sqrt{-g}\left\{\frac{1}{2\kappa^2}R
 - \frac{1}{2}\partial_\mu \varphi \partial^\mu \varphi
 +\frac{1}{2}\left(\frac{2 c_2 \alpha  (t_s-\chi )^{-1+\alpha }}{\kappa ^2}\right)\partial_\mu \chi\partial^\mu \chi
 - \tilde{V}_{\chi}(\varphi,\chi)\right\}\, .
\ee
with $\tilde{V}_{\chi}(\varphi,\chi)$ being equal to,
\begin{equation}\label{newtildepot}
\tilde{V}_{\chi}(\varphi,\chi)\simeq C_0-\frac{c_5 \alpha  (t_s-\chi )^{-1+\alpha }}{\kappa ^2}+\frac{6 c_4 c_5 (t_s-\chi )^{\alpha }}{\kappa ^2}+\frac{3 c_5^2 (t_s-\chi )^{2 \alpha }}{\kappa ^2}
\end{equation}
where we neglected the third term in the second line of Eq. (\ref{finalpotential}), since it becomes exponentially suppressed in comparison to the other $\chi$-containing terms. This late-time behavior has interesting consequences in the vacuum theory when matter fluids are not present. This will be the subject of a later section.

\subsubsection{Numerical analysis of the evolution of the scalar field $\chi$}

In this section we analyze numerically the evolution of the scalar $\chi$. Recall that we want to achieve an evolution for which the contribution of the scalar field $\chi$ is negligible before, during and after the inflationary era, and dominates at late-time. In order to achieve this, we shall assume that the evolution of the scalar field $\chi$, does not satisfy the slow-roll conditions. It is worth recalling the slow-roll conditions for a canonical scalar field $\sigma$. For a detailed presentation with regards to these issues, the reader is referred to Refs.~\cite{inflation,inflationreview}. In the context of the slow-roll condition, the following constraint is assumed to hold true,
\begin{equation}\label{slowrollconstr}
\frac{1}{2}\dot{\sigma}^2\ll V(\sigma)\, ,
\end{equation}
and in addition it assumed that this constraint is valid for an extended period of time. The constraint (\ref{slowrollconstr}) is known as the first slow-roll condition, and it ensures a long and finite acceleration period. In order the constraint (\ref{slowrollconstr}) is valid for a large period of time, the following additional constraint must be imposed,
\begin{equation}\label{slowrollconstrnew}
|\ddot{\sigma}|\ll \left| \frac{\partial V(\sigma)}{\partial \sigma} \right|\, ,
\end{equation}
which quantifies a constraint known as the second slow-roll condition. The canonical scalar field equation of motion in a FRW background is equal to,
\begin{equation}\label{asxeto}
\ddot{\sigma}+3H\dot{\sigma}+V'(\sigma)=0
\end{equation}
with the prime denoting differentiation with respect to $\sigma$. In virtue of Eq. (\ref{asxeto}), the constraint (\ref{slowrollconstrnew}) can be rewritten in the following form,
\begin{equation}\label{newcondqweeee}
|\ddot{\sigma}|\ll 3H |\dot{\sigma}| \, .
\end{equation}
By combining the two slow-roll conditions (\ref{slowrollconstr}) and (\ref{slowrollconstrnew}), the equation of motion of a canonical scalar field $\varphi$ in the slow-roll approximation becomes,
\begin{equation}\label{eqnsmotoinscla}
\dot{\sigma }\simeq -\frac{1}{3H}\frac{\partial V(\sigma )}{\partial \sigma} \, ,
\end{equation} 
In the case that the canonical scalar field is transformed to a non-canonical scalar field $\chi$, via the relation,
\begin{equation}\label{noncanchi}
\sigma =\int^{\chi}\sqrt{\eta (\chi)}\mathrm{d}\chi
\end{equation}
then, the slow-roll equation of motion for the non-canonical scalar field $\chi$, can be cast in the following way,
\begin{equation}\label{noncanonicalslowroll}
3H\eta (\chi) \dot{\chi}+V'(\chi)=0
\end{equation}
where this time, the prime indicates differentiation with respect to the non-canonical scalar field $\chi$. In addition, the non- slow-roll equation of motion of the scalar field $\chi$ is given in Eq. (\ref{A9}), but we quote it here again for convenience,
\begin{equation}\label{nonslowrolleqnmotion}
\eta(\chi)\ddot\chi + \frac{1}{2}\eta'(\chi) {\dot \chi}^2
+ 3H\eta(\chi)\dot\chi + \frac{\partial \tilde V(\phi,\chi)}{\partial 
\chi}=0 
\end{equation}
We shall solve this equation numerically, in order to see if our argument for the evolution of  the non-canonical scalar field $\chi$, remains valid. We choose the parameters $c_i$, $i=1,..5$ in the following way,
\begin{equation}\label{parameters}
c_1=\frac{3}{4},{\,}{\,}{\,}c_2=10^{-40},{\,}{\,}{\,}c_3=10^{-28},{\,}{\,}{\,}c_4=-\frac{3}{4},{\,}{\,}{\,}c_5=10^{-38}
\end{equation}
So these values of the parameters satisfy the assumptions we made previously in Eqs. (\ref{constrmodel1}) and (\ref{ndhdgf}). In addition, we shall take into account that the present time in seconds is approximately $t_p\simeq 4.25\times 10^{17}$sec and the constant $\kappa$ is $\kappa=8\pi G=2.0944\times 10^{-18}$GeV$^{-1}$, with $G$ being Newton's constant. In addition, since we assumed $\alpha>1$, we chose $\alpha=4/3$ and also that $t_s=10^{-35}$sec, the cosmic time that inflation approximately ended. There is another important reason behind that choice of $\alpha$, since when $\alpha$ is of the form given in Eq. (\ref{IV2}), with $n$ an even integer, and larger than one, the scalar field $\chi$ is always a non-phantom scalar. Of course this strongly depends on the initial conditions and the values of the rest parameters, but this is crucially determined by the choice $n=$even. We shall discuss later on what happens in the case $\alpha>1$ and $\alpha=n/(2m+1)$, with $n=$odd integer. For the choices of the parameters we made in Eq. (\ref{parameters}), it follows that the second scalar $\chi$ is non-phantom for all times $t$.

\begin{figure}[h]
\centering
\includegraphics[width=20pc]{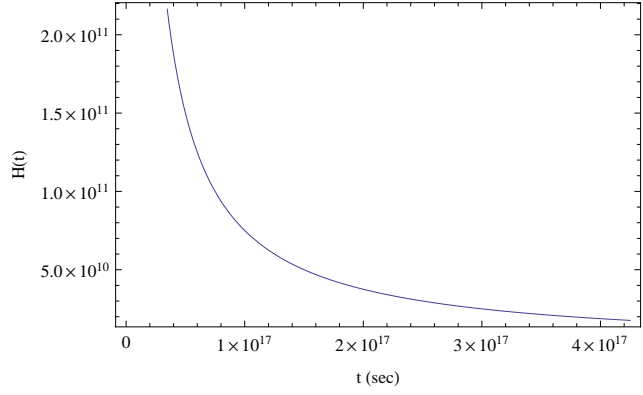}
\caption{\it{{The Hubble rate $H(t)=\frac{c_1}{c_2+c_3 t}+c_4+ c_5 \left( t_s - t \right)^\alpha$ as a function of the cosmic time with $c_1=\frac{3}{4}$, $c_2=10^{-40}$, $c_3=10^{-28}$, $c_4=-\frac{3}{4}$, $c_5=10^{-38}$, $t_s=10^{-35}$sec and $\alpha=4/3$.}}}
\label{plot1}
\end{figure}

\begin{figure}[h]
\centering
\includegraphics[width=20pc]{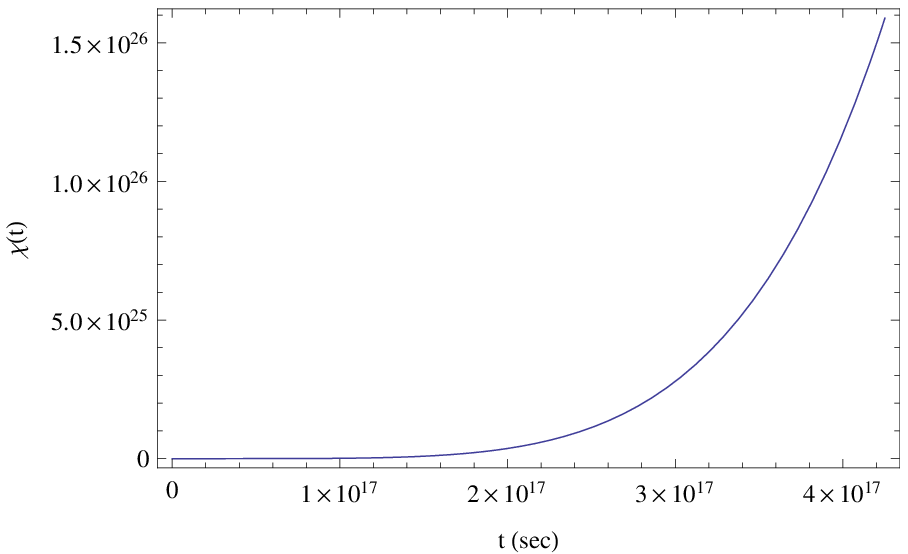}
\caption{\it{{The scalar field's $\chi (t)$ evolution as a function of cosmic time $t$ with $c_1=\frac{3}{4}$, $c_2=10^{-40}$, $c_3=10^{-28}$, $c_4=-\frac{3}{4}$, $c_5=10^{-38}$, $t_s=10^{-35}$sec, $\chi (10^{-40})\simeq 10^{-20}$, $\chi '(10^{-40})\simeq 10$ and $\alpha=4/3$. }}}
\label{plot3}
\end{figure}

\begin{figure}[h]
\centering
\includegraphics[width=15pc]{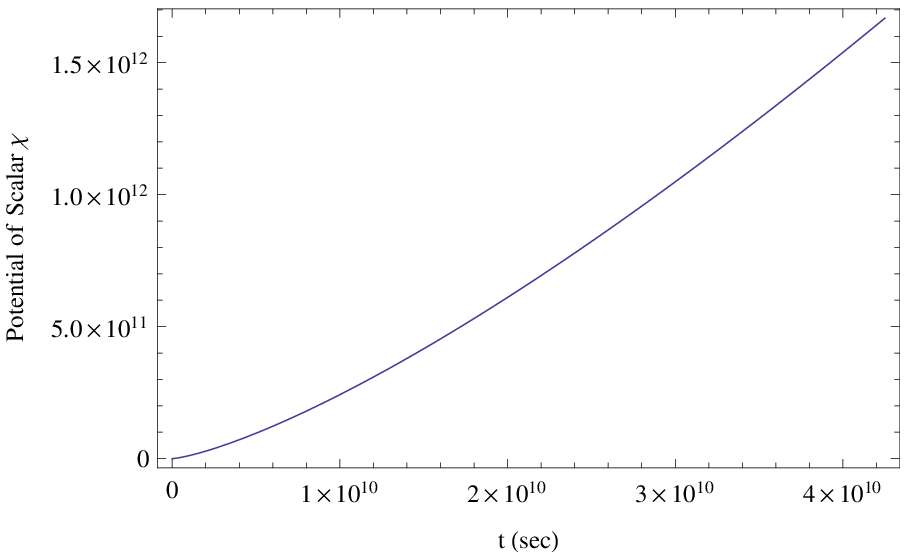}
\caption{The scalar field's $\chi (t)$ scalar potential $\tilde{V}_{\chi}$ as a function of cosmic time $t$ with $c_1=\frac{3}{4}$, $c_2=10^{-40}$, $c_3=10^{-28}$, $c_4=-\frac{3}{4}$, $c_5=10^{-38}$, $t_s=10^{-35}$sec, $\chi (10^{-40})\simeq 10^{-20}$ and $\chi '(10^{-40})\simeq 10$}\label{plot8}
\end{figure}

Using these values for the parameters, in Fig. \ref{plot1}, we have plotted the behavior of the Hubble rate as a function of the cosmic time $t$. As it can be seen, the Hubble rate is always positive with the choice of parameters we made. 
Having ensured that the Hubble rate takes only positive values, we proceed to solve numerically Eq. (\ref{nonslowrolleqnmotion}), in order to see if our assumption that the scalar field $\chi$ evolves from small values during inflation, to larger values at late-time where it dominates the potential. 
\begin{figure}[h]
\centering
\includegraphics[width=20pc]{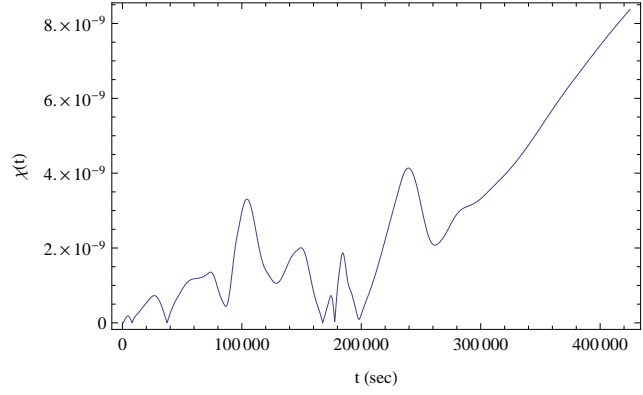}
\caption{\it{{The scalar field's $\chi (t)$ evolution as a function of cosmic time $t$ with $c_1=\frac{3}{4}$, $c_2=10^{-40}$, $c_3=10^{-28}$, $c_4=-\frac{3}{4}$, $c_5=10^{-38}$, $t_s=10^{-35}$sec, $\chi (10^{-40})\simeq 10^{-20}$, $\chi '(10^{-40})\simeq 10$ and $\alpha=4/3$. }}}
\label{plot2}
\end{figure}
We assume that the initial values at $t=10^{-40}$sec (approximately before inflation starts) for the scalar field $\chi$ are $\chi (10^{-40})\simeq 10^{-20}$ and $\chi '(10^{-40})\simeq 10$. As we can see in Fig. (\ref{plot2}), the scalar field $\chi$ remains small even for $t\simeq 10^{30}$. In Fig. (\ref{plot3}), it can be seen that after $t\simeq 10^{38}$sec, the scalar field $\chi $ grows larger and larger.
In addition, let us calculate the value of the scalar potential (\ref{newtildepot}) for $t=10^{-35}$sec. We assumed that the potential of the second scalar field $\varphi$ dominates at that time so we explicitly verify this here. Indeed, the value of the scalar field $\chi$ at $t=10^{-35}$sec is, $\chi (10^{-35})=10^{-20}$ and for this, the potential (\ref{newtildepot}) is approximately equal to $\tilde{V}_{\chi}(\varphi,\chi)\simeq 10^{-17}$. Notice that we assumed that the scalar field $\varphi$ near $t=10^{-35}$sec takes quite large values and particularly those adopted in \cite{sergeistarobinsky}, so that all the exponentials of the scalar field $\varphi$ are approximately equal to one. At the same time, for large values of $\varphi$ and with the constants $c_i$ chosen as in (\ref{parameters}), the scalar potential of the scalar field $\varphi$, given in Eq. (\ref{vsnew}), is approximately equal to $V_s(\varphi)\simeq 18.75 \times 10^{6}$, so our approximation is valid. 
 \begin{figure}[h]
\centering
\includegraphics[width=15pc]{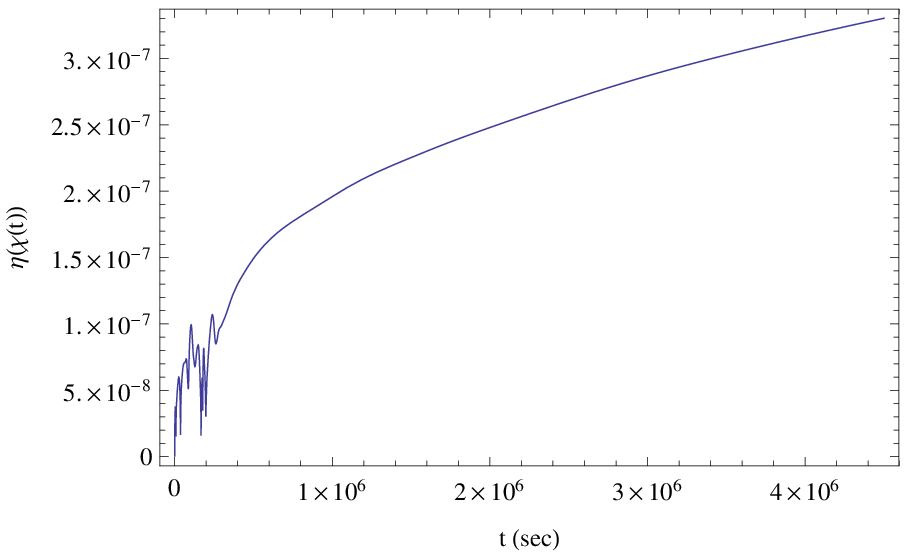}
\includegraphics[width=15pc]{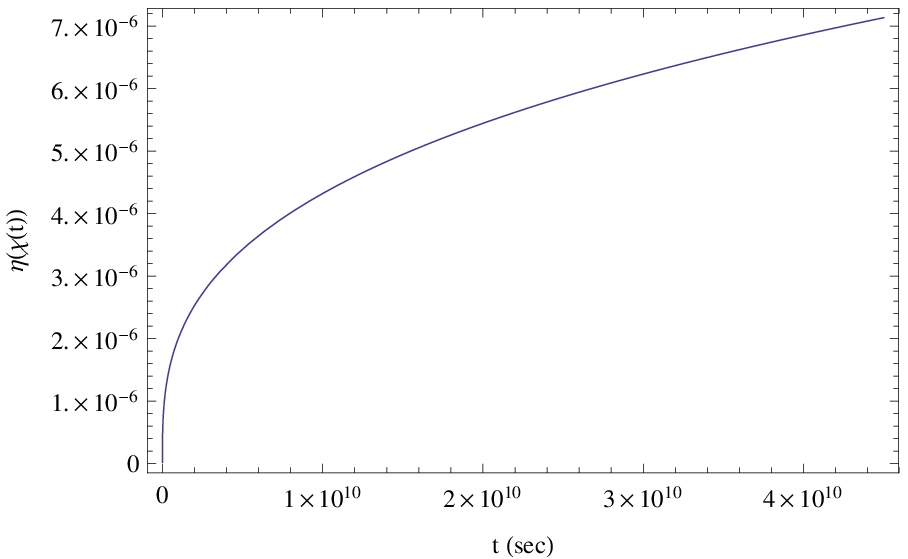}
\caption{The scalar field's $\chi (t)$ kinetic function $\eta (\chi (t))$ as a function of cosmic time $t$ with $c_1=\frac{3}{4}$, $c_2=10^{-40}$, $c_3=10^{-28}$, $c_4=-\frac{3}{4}$, $c_5=10^{-38}$, $t_s=10^{-35}$sec, $\chi (10^{-40})\simeq 10^{-20}$, $\chi '(10^{-40})\simeq 10$ and $\alpha=4/3$.}
\label{plot4}
\end{figure}
A very important comment is in order. The evolution of the scalar field $\chi$ behaves in the way we just described, only if it is assumed that it's evolution does not follow the slow-roll approximation. Indeed, if we assume that the field $\chi$ satisfies the slow-roll conditions, then by solving numerically the slow-roll equation of motion for the scalar $\chi$, namely Eq. (\ref{noncanonicalslowroll}), for the potential (\ref{newtildepot}), we may obtain the value of the scalar field $\chi(t)$, at $t=10^{-35}$sec, which is $\chi \sim 10^{10}$, so the potential is approximately equal to $\tilde{V}_{\chi}(\varphi,\chi)\sim 10^{11}$, so our approximation is not valid. 
 \begin{figure}[h]
\centering
\includegraphics[width=15pc]{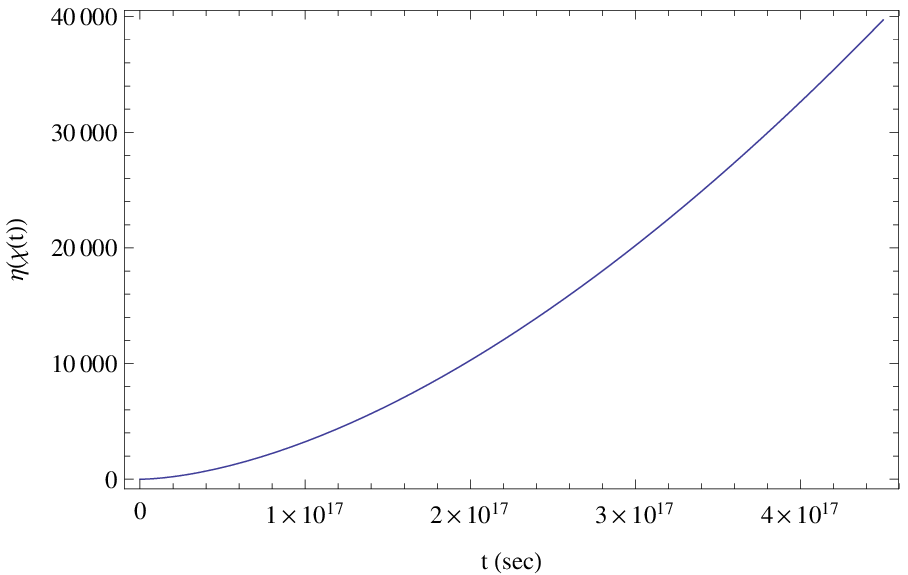}
\caption{The scalar field's $\chi (t)$ kinetic  function $\eta (\chi (t))$, as a function of cosmic time $t$ with $c_1=\frac{3}{4}$, $c_2=10^{-40}$, $c_3=10^{-28}$, $c_4=-\frac{3}{4}$, $c_5=10^{-38}$, $t_s=10^{-35}$sec, $\chi (10^{-40})\simeq 10^{-20}$, $\chi '(10^{-40})\simeq 10$ and $\alpha=4/3$.}
\label{plot5}
\end{figure}
Before closing this section, let us explicitly see that with the choices of the parameters and the initial conditions of the scalar field $\chi$, the scalar field $\chi$ is always non-phantom. Indeed, in Figs. \ref{plot4} and \ref{plot5} we have plotted the behavior of the function $\eta (\chi (t))$ for various time ranges. The first observation we need to make is that the scalar field's $\chi$ kinetic term always positive. This result crucially depends on the initial conditions and also to the form of $\alpha=n/(2m+1)$, with $n=$even. The second observation has to do with the values of the kinetic term. As can be seen in all figures, the kinetic function $\eta (\chi)$ is negligible, even for $t\simeq 10^6$sec, which corresponds to a time before the nucleosynthesis and during the era that the CMB spectrum is fixed. 
\begin{table}[h]
\begin{center}
\begin{tabular}{|c|c|c|c|c|}
  \hline
  Time & $t\simeq 10^{-34}$ &  $t\simeq 10^{-5}$ & $t\simeq 10^{5}$ & $t\simeq 10^{17}$ \\
  \hline
  $\eta (\chi (t) )$ & $1.43\times 10^{-11}$ & $3.8\times 10^{-11}$ & $9.5\times 10^{-8}$ & $6.97\times 10^{6}$ \\
  \hline 
\end{tabular}
\end{center}
\caption{\label{kineteictable}The scalar field's $\chi (t)$ kinetic function $\eta (\chi (t))$ as a function of cosmic time $t$}
\end{table}
In addition, by looking at Table \ref{kineteictable}, we can see that $\eta (\chi)$ remains indeed small for a long period of time. Of course we should bear in mind that the model we present is just a toy model, and a more concrete analysis should be performed, including matter fluids. Nevertheless, our assumption that the scalar $\chi$ makes negligible contribution during the inflationary era and for a long time period after is true. 
 \begin{figure}[h]
\centering
\includegraphics[width=15pc]{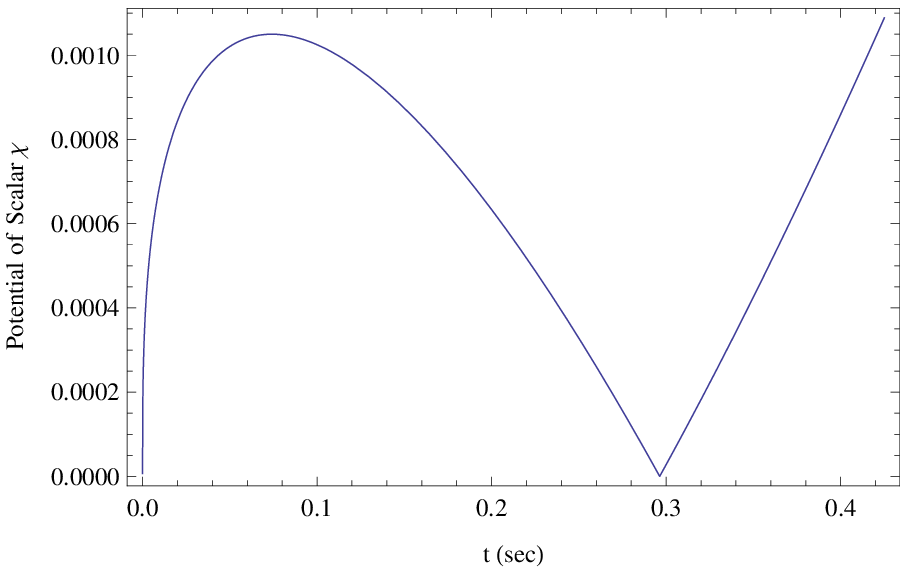}
\includegraphics[width=15pc]{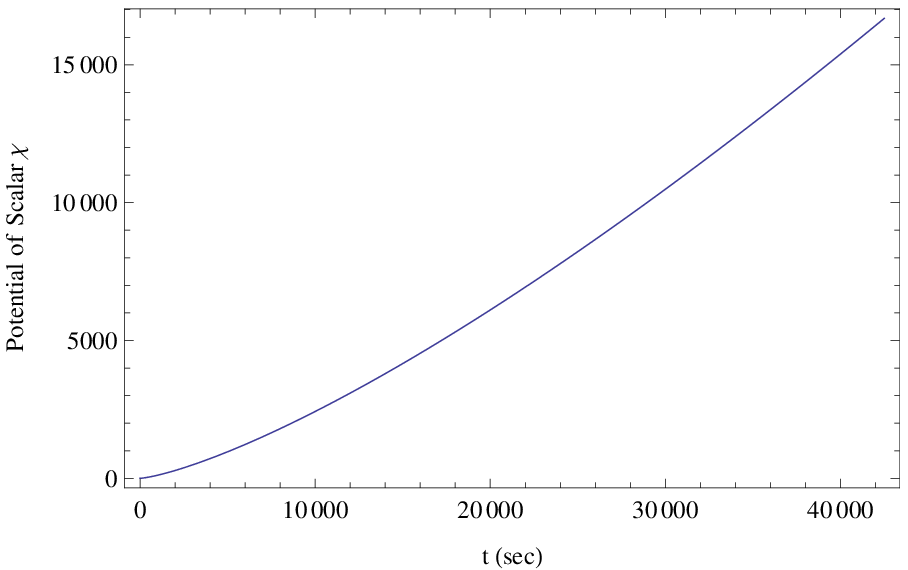}
\caption{The scalar field's $\chi (t)$ scalar potential $\tilde{V}_{\chi}$ as a function of cosmic time $t$ with $c_1=\frac{3}{4}$, $c_2=10^{-40}$, $c_3=10^{-28}$, $c_4=-\frac{3}{4}$, $c_5=10^{-38}$, $t_s=10^{-35}$, $\chi (10^{-40})\simeq 10^{-20}$, $\chi '(10^{-40})\simeq 10$ and $\alpha=4/3$.}
\label{plot7}
\end{figure}
In order to further support this, in Figs. \ref{plot7} and \ref{plot8}, we have plotted the behavior of the scalar potential of $\chi$ as a function of time. As it can be seen, the potential is quite small for a large period of time and also it is negligible during the inflationary era. 
In addition, this behavior can also be verified by looking at Table \ref{potentialtable}, where it can be seen that the potential of the scalar field $\chi$ is negligible for a long period of time, and starts to be significant at present or much after than present time.
\begin{table}[h]
\begin{center}
\begin{tabular}{|c|c|c|c|c|}
  \hline
  Time & $t\simeq 10^{-34}$ &  $t\simeq 10^{-7}$ & $t\simeq 10^{7}$ & $t\simeq 10^{17}$ \\
  \hline
  $\tilde{V}_{\chi}$ & $7.46\times 10^{-15}$ & $7.76\times 10^{-6}$ & $1.2\times 10^{7}$ & $2.6\times 10^{20}$ \\
  \hline 
\end{tabular}
\end{center}
\caption{\label{potentialtable}The scalar field's $\chi (t)$ scalar potential $\tilde{V}_{\chi}$ as a function of cosmic time $t$}
\end{table}

\begin{figure}[h]
\centering
\includegraphics[width=15pc]{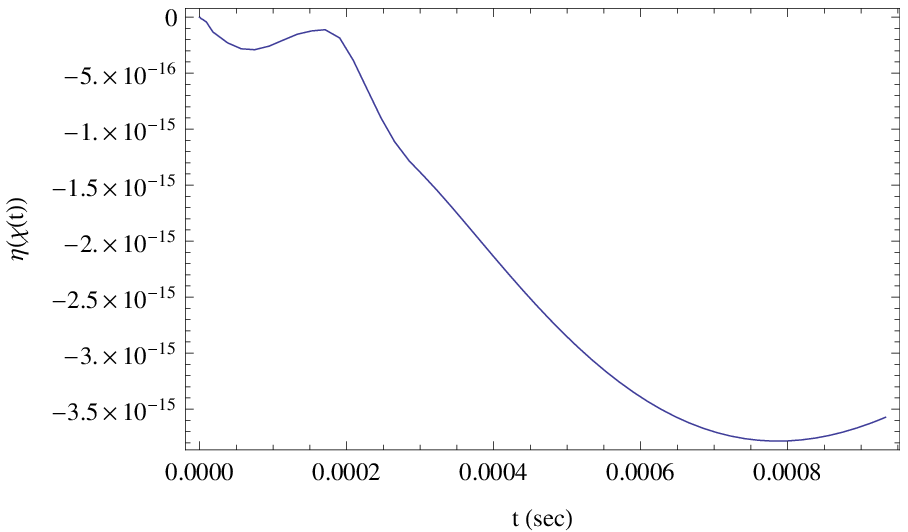}
\caption{The scalar field's $\chi (t)$ kinetic function $\eta (\chi (t))$ as a function of cosmic time $t$ with $c_1=\frac{3}{4}$, $c_2=10^{-40}$, $c_3=10^{-28}$, $c_4=-\frac{3}{4}$, $c_5=10^{-38}$, $t_s=10^{-35}$sec, $\chi (10^{-40})\simeq 10^{-20}$ and $\chi '(10^{-40})\simeq 10$, for $\alpha=5/3$}\label{plot12}
\end{figure}

Before closing this section, we shall verify that indeed the value of the parameter $\alpha$ appearing in the Hubble rate of the model, given in Eq. (\ref{IV1Bnoj}), plays a crucial role in determining the phantom- non-phantom behavior of the scalar field $\chi$. Indeed, if instead of choosing $\alpha=n/(2m+1)$, with $n=$even, we choose, $n=$odd, then the scalar field $\chi$ is always a phantom scalar. This can also be seen in Fig. \ref{plot12}, where we plotted the kinetic function $\eta (\chi (t))$ as a function of cosmic time $t$, for $\alpha=5/3$.

Finally, let us note that the parameter space we used seems very much constrained, but in principle the same qualitative behavior appears for a wide range of the parameters. The fine-tuning imposed in the parameter space was necessary so that the slow-roll dynamics of the canonical scalar field $\varphi$ governs the early-time dynamics and so the singularity does not affect the early-time dynamics, since it governs the evolution of the second scalar field $\chi$. Notice that the slow-roll approximation condition is very necessary for this resulting dynamical evolution. We need to note that when we say for a wide range of parameters we mean that $c_2$ for example instead of being $c_2=10^{-40}$, it can be chosen to be $c_2=10^{-30}$ and $c_3$, instead of being $c_3=10^{-28}$ it can be chosen $c_3=10^{-18}$. So practically the parameters can take a continuum of values, so long as the fractions between $c_2/c_3$, $c_2/c_5$ and $c_3/c_5$ remain constant. So by fine tuning we mean exactly this and not fixing the values of the parameters to take a specific value.

\subsection{Slow-roll Parameters and observational indices of the $R^2+R+\Lambda$ model}

Having presented the essentials of the canonical scalar field model with potential which at early times is given by Eq. (\ref{vsnew}), in this section we shall extensively study the implications of this model on the slow-roll parameters and correspondingly on the observational indices. This analysis was also performed for a similar model in Ref. \cite{sergeistarobinsky}, so the reader is referred to \cite{sergeistarobinsky} for details. The canonical scalar theory with potential (\ref{vsnew}) corresponds to an $R+R^2$ gravity plus a cosmological constant in the Jordan frame, of the form given in Eq. (\ref{pure}). By the form of the potential and for the values of the parameters (\ref{parameters}), the potential (\ref{vsnew}) can be further simplified to the following form,
\begin{equation}\label{vsnewobserv}
V_{s}(\varphi)\simeq C_0+Ae^{-2\sqrt{\frac{2}{3}}\kappa   \varphi }-2 Ae^{-\sqrt{\frac{2}{3}}\kappa   \varphi }
\end{equation}
with $A=\frac{3c_1^2}{\kappa^2}$, since $c_1=-c_4$. Notice that we disregarded the term $\frac{c_1 c_3 e^{-2\sqrt{\frac{2}{3}}\kappa   \varphi }}{\kappa ^2 }$, because for the values of the parameters as in (\ref{parameters}), it is subdominant compared to the other two terms. The minimum of the potential is at $\varphi=0$, since the equation $V'(\varphi)=0$ yields the solution $\varphi=0$, for which the second derivative is equal to $V''(0)=\frac{4 A^2\kappa^2}{3}$, which is positive, so the critical point $\varphi=0$ is a global minimum of the potential $V_s(\varphi )$. Following \cite{sergeistarobinsky}, and by assuming slow-roll evolution for the scalar field $\varphi$, the slow-roll equations for $\varphi$ during inflation read,
\begin{equation}\label{slowrollduringinflation}
H^2\simeq \frac{\gamma}{12},{\,}{\,}{\,}3H\dot{\varphi}\simeq \left(\frac{\gamma}{\sqrt{6\kappa^2}}\right)e^{-\sqrt{\frac{2}{3}}\kappa   \varphi }
\end{equation}
with $\gamma=\frac{9}{4}$. From Eq. (\ref{slowrollduringinflation}) it follows that quasi de Sitter acceleration is realized. The scalar field behaves as follows,
\begin{equation}\label{solutionofscalarfields}
\varphi \simeq \sqrt{\frac{2}{3}\kappa^2} \ln \left ( \frac{1}{3}\sqrt{\frac{2\gamma}{3}}\left(t_0-t\right)\right)
\end{equation}
with $t_0$ bounded at the beginning of inflation. Since during that time, the scalar field's $\varphi$ values are quite large, the slow-roll parameters are quite small, so that the scalar field's evolution proceeds slowly. Inflation ends when the slow-parameters become of order one and in order that we obtain $N=60$ e-foldings we must require that the initial value of the scalar field is approximately equal to $\varphi_i\simeq \frac{1.07}{\kappa^2} $, which is quite large (see \cite{sergeistarobinsky} for details). For this model, the primordial power spectrum is,
\begin{equation}\label{spectrum}
\Delta_R^2\simeq\frac{\kappa^2\gamma^2N^2}{72\pi^2},
\end{equation}
while the spectral index $n_s$ and the scalar-to-tensor ratio $r$ as a function of the e-folding number $N$ read,
\begin{equation}\label{spectrobserv}
n_s\simeq 1-\frac{2}{N},{\,}{\,}r\simeq \frac{12}{N^2}
\end{equation}
Therefore, for $N=60$, we have $n_s\simeq 0.9665$, and also $r\simeq 0.0029$, which are compatible with the Planck data \cite{planck}. Consequently, the behavior of the model with scalar potential (\ref{vsnew}) is almost identical to the $R^2$ inflation model, with the only difference being the fact that the minimum of the potential is shifted in our case, since $V_s(0)\simeq C_0-A$. In addition, when the corresponding Jordan frame $F(R)$ gravity is considered, the difference with the $R^2$ inflation theory, is the appearance of a cosmological constant $\Lambda$, defined in Eq. (\ref{pure}). As was also pointed out in \cite{sergeistarobinsky}, the appearance of this cosmological constant at large curvatures needs to be explained, since it may originate from possible quantum effects. Regardless of that, the presence of a cosmological constant does not modify the inflationary properties of the canonical scalar's field potential.

We need to note here that the model we study in this paper, which consists of two scalar fields, and the single scalar field model we studied in Ref. \cite{Nojiri:2015fra}, have some qualitative differences. Indeed, the present two scalar field model is constructed in such a way so that the singular evolution is governed by the second scalar field $\chi$, while the inflationary era is governed by the canonical scalar field which has a nearly $R^2$ potential. Therefore, in some sense the singularity is hidden in the $\chi$ sector and does not appear or does not affect the slow-roll evolution of the first scalar field $\varphi$. This observation is crucial, since without the assumption of the slow-roll approximation for the canonical scalar $\varphi$, the slow-roll parameters would not be given by the following equations,
\begin{equation}\label{slowrolpparaenedefs}
\epsilon =\frac{1}{2\kappa^2}\left(\frac{V '(\varphi )}{V (\varphi
)}\right),{\,}{\,}{\,}\eta
=\frac{1}{\kappa^2}\left(\frac{V''(\varphi )}{V (\varphi )}\right)
\, ,
\end{equation} 
but in contrast, from the following equations, which are known to describe the so-called Hubble slow-roll parameters \cite{barrowslowroll},
\begin{equation}\label{hubbleslowrolloriginal}
\epsilon_H =-\frac{\dot{H}}{H^2},{\,}{\,}{\,}\eta_H
=-\frac{\ddot{H}}{2H\dot{H}} \, .
\end{equation}
By the analytic form of the Hubble slow-roll parameters (\ref{hubbleslowrolloriginal})It is conceivable that without the slow-roll assumption for the canonical scalar field, the dynamics of the canonical scalar field would be strongly affected by the singularity, since for specific values of $\alpha$, the second slow-roll parameter would diverge. This could in principle have either catastrophic consequences for the inflationary era, if for example the singularity is chosen to occur before the end of the inflationary era, or lead to alternative physical phenomena, being related to the instability in the slow-roll parameters at the Type IV singularity. We shall not go proceed further towards this research line since this is out of the scope of the present paper, but a physical application of these effects is in progress. Finally, let us just comment on the possibility of having catastrophic consequences in the slow-roll indices, that in \cite{Nojiri:2015fra}, this was the case if the singularity was chosen to occur before the end of the inflationary era, even in the context of the slow-roll approximation. However in our case, the slow-roll approximation and the fine-tuning of the parameters, protect the two-scalar field system from such instabilities, at least at the level of the slow-roll parameters.

\subsubsection{Models II: Jordan frame $R^{\frac{n+2}{n+1}}$ and $R+C_1R^2+C_2(R+R_0)^{3/2}$ gravity}

The potential (\ref{vsnew}) offers much freedom with regards to the choice of the parameters $C_i$, $i=1,2,3$. Following the lines of research of the previous section, we choose the parameters $C_i$ to satisfy,
\begin{equation}\label{CIPARAME}
C_0=-C_1,{\,}{\,}{\,}C_2=-\frac{C_0}{4}
\end{equation}
Notice that $c_1=3/4$ in all cases. The choice (\ref{CIPARAME}), would imply for the parameters $c_i$, $i=2,...4$ the following,
\begin{equation}\label{thefolow}
c_4=-2c_1,{\,}{\,}{\,}c_3=6c_1
\end{equation}
With this choice of parameters, the potential (\ref{vsnew}), becomes during and at the end of the inflationary era,
\begin{equation}\label{newpotential}
\tilde{V}(\varphi,\chi)\simeq -\frac{\gamma (n+2)}{\kappa^2}+\frac{\gamma }{\kappa^2}e^{-2\sqrt{\frac{2}{3}}\kappa   \varphi }+\frac{\gamma (n+2)}{\kappa^2}e^{-\sqrt{\frac{2}{3}}\kappa   \varphi }
\end{equation}
where $\gamma$ an arbitrary parameter. Using Eqs. (\ref{solvequation}), (\ref{can}) and (\ref{newpotential}), we may obtain the Jordan frame $F(R)$ gravity which gives rise to the potential (\ref{newpotential}), which when $R\gg \gamma$, is approximately equal to,
\begin{equation}\label{largeeqnts}
F(R)\simeq \frac{3}{4}\frac{1}{16^{1/3}}R^{4/3}
\end{equation}
which may be further approximated $F(R)\sim R^{1.33}$.

By the same token, if we choose the parameters $c_i$, $i=1,..4$ to be,
\begin{equation}\label{newvaluesparameters}
c_1=3,{\,}{\,}{\,}c_3=\frac{c_1}{3},{\,}{\,}{\,}c_4=-\frac{3}{2}c_1
\end{equation}
and by defining $\gamma=9c_1^2$, then the corresponding Jordan frame gravity easily follows,
\begin{equation}\label{finaljordanframegravity}
F(R)=R+\frac{R^2}{6\gamma}+\frac{\sqrt{3}}{36}\left(\frac{4R}{\gamma}+3\right)^{3/2}+\frac{\gamma}{4}
\end{equation}

\subsection{A brief comment on Higgs inflation}

Before closing this section, we have to mention that the $R^2$ inflation potential (\ref{astarobisnkypotential}) has a direct correspondence to a certain limit of Higgs inflation model introduced by Bezrukov and Shaposhnikov \cite{shaposhnikov}. So practically, the possibility of having a singular evolution in the context of the $R^2$ inflation model, can be done in the Higgs inflation case too, at least at a certain limit. Indeed, the Jordan frame action of the Higgs inflaton model is (disregarding matter),
\begin{equation}\label{higgsaction}
S=\int \mathrm{d}x^4\sqrt{-g}\left[\frac{1}{2\kappa^2}R\left(1+2\xi\kappa^2H^{\dag}H\right)\right]
\end{equation}
with $H$ the Higgs doublet, which is equal to,
\begin{equation}\label{higgsdoublet}
H=\frac{1}{\sqrt{2}}\left(%
\begin{array}{c}
  0\\
  h \\
\end{array}%
\right)
\end{equation}
By performing a conformal transformation to the Einstein frame and by canonically normalizing the scalar field, in the limit where $h\gg 1$, the Higgs inflaton action can be cast in the following form,
\begin{equation}\label{higgsaction1}
S=\int \mathrm{d}x^4\sqrt{-g} \left(\frac{1}{2\kappa^2}R-\frac{1}{2}\partial_{\mu}\bar{h}\partial^{\nu}\bar{h}-\frac{\lambda }{\kappa^44\xi^2}\left(1-e^{-\sqrt{\frac{2}{3}}\kappa \bar{h}}\right)^2\right),
\end{equation}
where the canonical scalar field $\bar{h}$ is related to the scalar field $h$ as follows,
\begin{equation}\label{EHDG}
\bar{h}=\sqrt{\frac{2}{3\kappa^2}}\ln \left(1+\xi h^2\kappa^2\right).
\end{equation}
The analysis of the limiting case of the Higgs inflaton theory given in Eq. (\ref{higgsaction1}), can in principle proceed as in the $R^2$ inflation case, which we described earlier, so we omit it for brevity. Note however, that the Higgs inflaton theory has the behavior of Eq. (\ref{higgsaction1}) only for large field values, so it is expected that in the context of the $R^2$ inflation, there is more freedom for cosmological reconstruction model building.

\section{Non-singular dark energy era driven by Type IV singular inflation}

In this section we shall study in detail the cosmological evolution of the Model I we presented in the previous section. Recall that the Hubble rate for this model has the form given in Eq. (\ref{IV1Bnoj}), with the parameters being constrained in the way these appear in Eq. (\ref{constrmodel1}). In Table \ref{TableI} we have summarized the details of this model for convenience.
\begin{table}[h]
\begin{center}
\resizebox{\columnwidth}{!}
{
\begin{tabular}{|c|c|c|}
\hline
Hubble Rate & Constraints on the parameters  & Singularity Type
\\\hline
$H(t) =  \frac{c_1}{c_2+c_3 t}+c_4+ c_5 \left( t_s - t \right)^\alpha$,  & $
c_1=-c_4=\frac{3}{4},{\,}{\,}{\,}c_2\ll 1,{\,}{\,}{\,}c_3\ll 1,{\,}{\,}{\,}c_5\ll 1,  \alpha= \frac{n}{2m + 1}$ & Type IV for $\alpha>1$, $n=$even \\
\hline
\hline
Jordan Frame $F(R)$ Gravity  & $F(R)=R+\frac{R^2}{4C_0}+\Lambda$ & \\
\hline
 \end{tabular}}
\end{center}
\caption{\label{TableI}Brief Description of Model I and it's cosmological evolution}
\end{table}
Recall that with the choice of parameters we made in Eq. (\ref{parameters}), and also with the choice of $\alpha=n/(2m+1)$, with $\alpha >1$ and $n=$even integer, the scalar field $\chi$ is never a phantom scalar. This however can change for $\alpha=n/(2m+1)$, $n=$odd, and for appropriate choice of the initial conditions, but we briefly study the implications of this case in a later section.

It is worth recapitulate the approximations we made, because we shall make extensive use of this model in this section. At first, the cosmic time $t_s$ is considered to be the time at which inflation ends. At this point the two scalar fields action may be approximated by the action (\ref{A1newnearinfl}), so the inflationary era is dominated by the canonical scalar field $\varphi$ with a scalar potential $V_s(\varphi)$ given in Eq. (\ref{vsnew}). As the cosmic evolution proceeds and at late-time, the scalar field $\chi$ starts to dominate the evolution, and the action becomes approximately equal to (\ref{A1newnew}), with the potential that governs the late-time evolution being $\tilde{V}_{\chi}(\varphi,\chi)$, appearing in Eq. (\ref{newtildepot}). This model offers a theoretical framework for a quite appealing evolutionary process. Particularly, as we now explicitly demonstrate, the model near the Type IV singularity is governed solely by the nearly $R^2$ inflation potential (\ref{vsnew}), with the effects of the scalar field $\chi$ being disregarded, since these are in effect subdominant during and at the end of the inflationary era. The field $\chi$ is assumed to have small values during that era, and we choose it's initial conditions in such a way in order to achieve this. Notice that the Universe has no matter fluids present except only these two scalar fields. As the Universe evolves in time, then the effects of the scalar field begin to dominate, and at late-time the evolution is solely determined by the field $\chi$. Therefore we achieved to incorporate a Type IV singularity in a the cosmological evolution of a nearly $R^2$ inflation model, with the, apparently appealing, side effect of achieving singular $R^2$ inflation model during and at the end of inflation, and perhaps nearly phantom late-time acceleration. Particularly, the late-time acceleration maybe be driven directly by the Type IV singularity occurring at the end of inflation. In order to see this explicitly, we shall study the effective equation of state (EoS) parameter $w_{eff}$,
\begin{equation}
\label{eos}
w_\mathrm{eff}=\frac{p}{\rho}=-1-\frac{2\dot{H}}{3H^2} \, .
\end{equation}
which for the Hubble rate (\ref{IV1Bnoj}) reads,
\begin{equation}\label{weeffnewstudy}
w_{\mathrm{eff}}=-1-\frac{2 \left(-\frac{c_1 c_3}{(c_2+c_3 t)^2}-c_5 (-t+t_s)^{-1+\alpha } \alpha \right)}{3 \left(c_4+\frac{c_1}{c_2+c_3 t}+c_5 (-t+t_s)^{\alpha }\right)^2}
\end{equation}
Near the Type IV singularity, which occurs at the end of inflation, the Type IV singularity behaves as follows,
\begin{equation}\label{weffneartypeiv}
w_{\mathrm{eff}}\simeq -1+\frac{2 (c_1 c_3)}{3 c_1^2}
\end{equation} 
where we took into account the constraints (\ref{constrmodel1}) and also that as $t\rightarrow t_s$,
\begin{equation}\label{tsapprox}
(-t+t_s)^{\alpha-1}\simeq 0,{\,}{\,}{\,}(-t+t_s)^{\alpha}\simeq 0
\end{equation}
As is obvious, Eq. (\ref{weffneartypeiv}) describes quintessential acceleration which occurs near the Type IV singularity, and is controlled by the nearly $R^2$ inflation potential (\ref{vsnew}). In conclusion we must emphasize the important result that inflation is not phantom but an almost de Sitter accelerating phase, with almost de Sitter meaning nearly quintessential acceleration.

\begin{figure}[h]
\centering
\includegraphics[width=15pc]{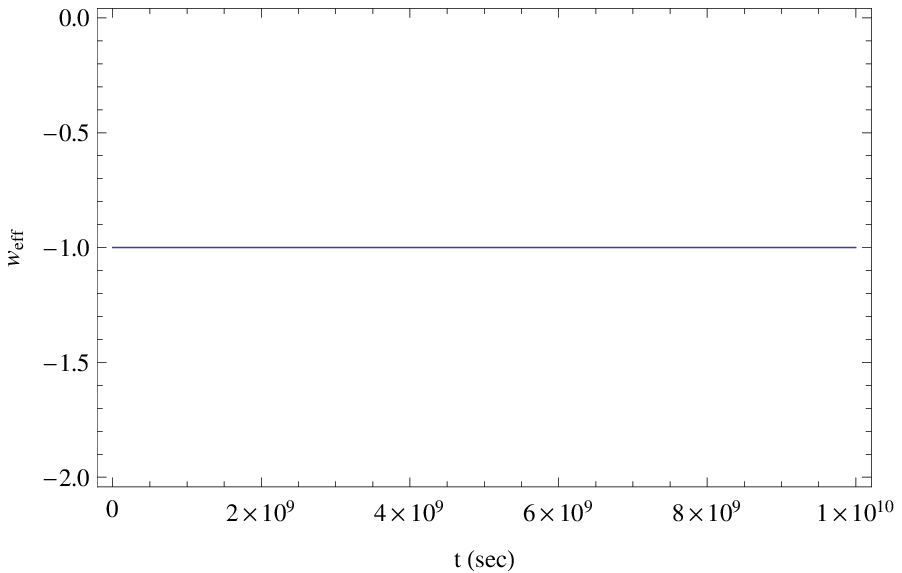}
\includegraphics[width=15pc]{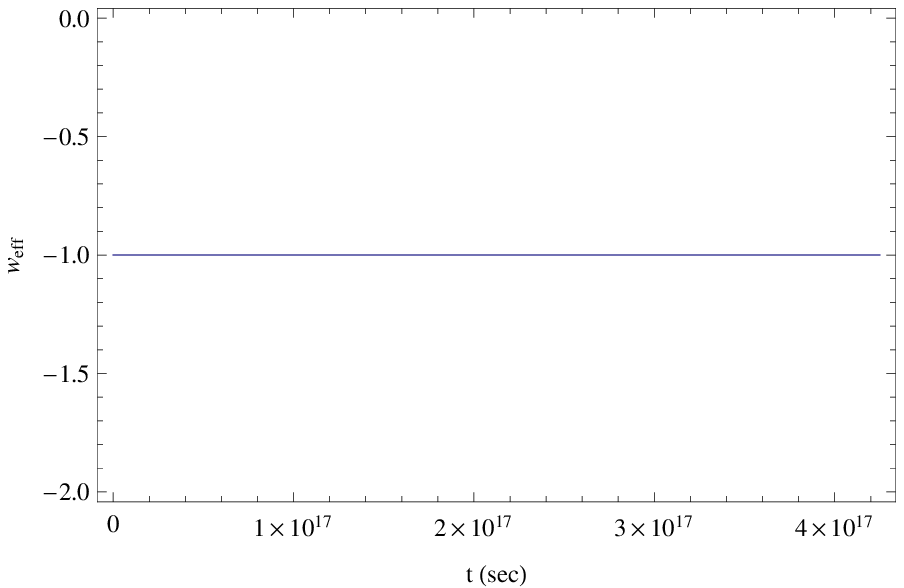}
\caption{The effective equation of state $w_{\mathrm{eff}}$ as a function of cosmic time $t$ with $c_1=\frac{3}{4}$, $c_2=10^{-40}$, $c_3=10^{-28}$, $c_4=-\frac{3}{4}$, $c_5=10^{-38}$, $t_s=10^{-35}$sec and $\alpha=4/3$}
\label{plot10}
\end{figure}

Correspondingly, at late-time the EoS takes the following form,
\begin{equation}\label{lateeos}
w_{\mathrm{eff}}\simeq -1-\frac{2 \left(-\frac{c_1 c_3}{(c_2+c_3 t)^2}+c_5 t^{-1+\alpha } \alpha \right)}{3 \left(c_4+\frac{c_1}{c_2+c_3 t}-c_5 t+^{\alpha }\right)^2}
\end{equation}
where we used the fact that, since $\alpha$ is of the form (\ref{IV2}), with $n=$even, the following holds true for large cosmic times,
\begin{equation}\label{largetaprrox}
(-t+t_s)^{\alpha-1}\simeq (-t)^{\alpha-1}=-t^{\alpha-1}
\end{equation}
since $\alpha-1$ is a fraction in general, with odd numerator. Relation (\ref{lateeos}) can describe phantom or non-phantom acceleration, depending on the choice of the parameters and the cosmic time value. With the choice of parameters we made in Eq. (\ref{parameters}), the effective equation of state (\ref{lateeos}), becomes, 
\begin{equation}\label{approxlateenqs}
w_{\mathrm{eff}}\simeq -1+\frac{0.8788\ 10^{-38} t^{1/3}}{\left(-0.75-1.33\ 10^{-38} t^{1/3}+\frac{0.75}{10^{-40}+10^{-28} t}\right)^2}-\frac{4.95\ 10^{-29}}{\left(10^{-40}+10^{-28} t\right)^2 \left(-0.75-1.33\ 10^{-38} t^{1/3}+\frac{0.75}{10^{-40}+10^{-28} t}\right)^2}
\end{equation}
From this equation, it is obvious that when $t\simeq t_p$, the first term of Eq. (\ref{approxlateenqs}), is approximately equal to $1.097 \times 10^{-54}$, while the second term is $8.8\times 10^{-29}$, so the second term dominates. It is obvious that the EoS is nearly phantom, since $w_{\mathrm{eff}}\simeq -1-8.8\times 10^{-29}$, but the phantom contribution is negligible, so practically the acceleration is almost a de Sitter one. The behavior of the EoS for times up to the present age of the Universe can be seen in Fig. \ref{plot10}, were it is obvious that the acceleration is practically de Sitter for all times. However, we need to stress that this behavior is strongly dependent on the choices of the parameters and also on the choice of $\alpha $. 
Recall that the late-time behavior for the two scalar model we just described, is governed by the scalar field $\chi$. In Table (\ref{tableII}) we gathered our results, with regards to the EoS behavior. So finally, as we demonstrated, the Type IV singularity that occurs at the end of the inflationary era, is responsible for the late-time acceleration. 
\begin{table}[h]
\begin{center}
\begin{tabular}{|c|c|}
\hline
EoS near the Type IV singularity & nearly Quintessential Acceleration
\\
\hline
EoS at late-time & Nearly Phantom Acceleration
 \\
\hline
Scalar potential near Type IV singularity & $V_{s}(\varphi )$, dominated by $\varphi$
 \\
\hline
Scalar potential at late-time & $\tilde{V}_{\chi}(\varphi,\chi)$, dominated by $\chi$
 \\
\hline
 \end{tabular}
\end{center}
\caption{\label{tableII}Summary of the cosmological evolution of Model I}
\end{table}
Finally we have to note that this picture entirely changes if the singularity is assumed to occur at late-time. Then the late-time behavior is no longer controlled by the Type IV singularity, but the effects of the singularity might affect the inflationary era, since $t_s$ might be chosen to be very big. We shall present the details of this scenario in the next section in the cases that the scalar field $\chi$ is non-phantom and phantom. Notice however that the impact of a phantom scalar on the scalar potential and also in the kinetic function $\eta (\chi)$ would completely alter the results we obtained in the previous sections, so the study of the phantom scalar is just included here for completeness. More emphasis was given in this section for the case of a non-phantom scalar and the singularity occurs at early time. 

Before going into this, we need to discuss another important issue, having to do with the value of the EoS parameter and the type of acceleration that the Universe experiences during inflation and at late-time. Particularly, near the Type IV singularity, which occurs at the end of inflation, the EoS is given in Eq. (\ref{weffneartypeiv}), which as we stated, describes nearly quintessential acceleration. However, with the use of the parameter values given in Eq. (\ref{parameters}), this quintessential acceleration is very close to de Sitter acceleration, since the second term of (\ref{weffneartypeiv}), namely, $\frac{2 (c_1 c_3)}{3 c_1^2}$, is approximately equal to, $2.66\times 10^{-28}$, which is extremely small. So nearly quintessential acceleration means that the acceleration is very close to de Sitter expansion. The same applies for the late-time behavior, as we showed earlier, but in this case the acceleration is nearly phantom, but very close to de Sitter. So practically, the acceleration in this model is always nearly de Sitter, but note that no matter fluids apart from the two scalars, were considered in this work.

Finally, let us discuss a vague point that may cause some confusion. In the analysis we performed we took the present time to be $t_p\simeq 10^{17}$sec. However, this estimate value should be calculated in our case by taking also into account the cold dark matter contribution to the total energy density. However, in our presentation we aimed in describing the features of the model qualitatively, at least with regards to the behavior of the EoS. The complete analysis should involve detailed numerical analysis, by taking into account all the observable numerical data, but such an analysis is beyond the scopes of this qualitative presentation.

\subsection{Other cases}

\subsubsection{Non-phantom field case}

In the case that $t_s\gtrsim t_p$, with $t_p$ denoting the present time, which is approximately $t_p\simeq 4.25\times 10^{17}$sec, the behavior of the resulting functional form of the EoS is altered drastically, since $t_s$ dominates over the term $t$. Indeed, when $t\simeq t_e$, with $t_e$ denoting the time that inflation ends, the effective EoS is approximately equal to,
\begin{equation}\label{wefflargets}
w_{\mathrm{eff}}\simeq -1-\frac{2 \left(-\frac{c_1 c_3}{(c_2+c_3 t)^2}-c_5 (t_s)^{-1+\alpha } \alpha \right)}{3 \left(c_4+\frac{c_1}{c_2+c_3 t}+c_5 (t_s)^{\alpha }\right)^2}
\end{equation}
However, if we choose the values of the parameters as in Eq. (\ref{parameters}) and also that $t_s\simeq 10^{50}$sec, the EoS becomes,
\begin{equation}\label{wefflargets1}
w_{\mathrm{eff}}\simeq -1-\frac{0.66 \left(-1.33 10^{-18}-\frac{7.5\ 10^{-19}}{\left(10^{-40}+10^{-28} t\right)^2}\right)}{\left(9.9\ 10^{41}+\frac{0.75}{10^{-40}+10^{-28} t}\right)^2}
\end{equation}
which means that for $t\simeq 10^{-35}$sec, the effective equation of state becomes approximately equal to,
\begin{equation}\label{wefflargets2}
w_{\mathrm{eff}}\simeq -1+8.82 \times 10^{-103}+4.97\times 10^{-23}
\end{equation}
which describes nearly quintessential acceleration, but almost de Sitter, since the terms that drive the quintessential acceleration are almost negligible. In the large $t$ regime, with $t_s\simeq 10^{50}$sec, when the cosmic time is nearly the present time, the functional form of the EoS is given by the same relation, namely Eq. (\ref{wefflargets}), and also for the values of the parameters given in Eq. (\ref{parameters}), the EoS becomes for $t\simeq 10^{17}$sec,
\begin{equation}\label{wefflargets3}
w_{\mathrm{eff}}\simeq -1+8.05 \times 10^{-103}+5.95\times 10^{-81}
\end{equation}
so the physical picture is the same as before, that is, nearly de Sitter but slightly quintessential acceleration. The same hold true if we choose $t_s\gg t_p$, with the terms that turn the EoS quintessential, being even more smaller.

An important remark is in order. If the time that the singularity occurs is chosen to be of the order of the present age of the Universe, or even larger, then the contribution of the second scalar field $\chi$ to the scalar potential at early times is not negligible anymore. A simple numerical calculation of the values of $\chi$, at $t\simeq 10^{-35}$sec with the same values of the parameters we used before and also the same initial conditions, yields $\chi (10^{-35})\simeq 10^{-20}$, and for this value, the potential of the scalar field $\chi$ is equal to $\tilde{V}_{\chi}\sim 10^{20}$, which destroys the physical picture we described earlier in this section. We will not go into further details towards this line of research, since the result is not so physical appealing.

\subsubsection{Phantom field case}

In order to study all possible cases, with regards to the behavior of the EoS, in this section we shall briefly discuss the case that the scalar field $\chi$ is always a phantom scalar. As we already mentioned earlier, this corresponds to the case that $\alpha=n/(2m+1)$, with $n=$odd integer and $\alpha>1$. The most interesting case corresponds to $t_s\simeq 10^{-35}$sec, which is the time when inflation ends. Suppose for example that $\alpha=5/3$, then the EoS at late-time is approximately equal to,
\begin{equation}\label{wefflargets5}
w_{\mathrm{eff}}\simeq -1+1.07 \times 10^{-97}+5.95\times 10^{-96}
\end{equation}
so the resulting picture is that, at times $t\simeq 10^{17}$sec, the EoS describes nearly quintessential, but practically de Sitter acceleration. We have to note that this result is owing to the fact that when $\alpha =5/3$, then $\alpha-1=2/3$, which renders the term $(-t+t_s)^{\alpha-1}$ robust towards the change of the sign of $(-t+t_s)$. Finally we need to stress that the physical picture might completely change if we choose other values of the parameters, so these results we presented hold true only for the choice of the parameters given in Eq. (\ref{parameters}).

\section{Stability analysis of the cosmological solution}

An attribute of using two scalar fields for the cosmological evolution, at least in the context of the reconstruction method we used, is that possible inconsistencies that may occur with the single scalar field description, do not usually occur in the case of two or more scalars. These instabilities were firstly observed in \cite{vikman,cai}, and they occur at exactly the transition point from non-phantom to phantom and vice-versa, in the context of single scalar field scalar-tensor cosmology. Usually, these instabilities are infinite instabilities \cite{Nojiri:2005pu,Capozziello:2005tf,vikman}, that is, a singularity might occur at the transition point. However, we must note that instabilities are not always an unwanted feature, in the presence of a Type IV singularity. We shall discuss this issue again in a later section. 

In this section we shall investigate the stability of the cosmological solution (\ref{A4}), with a Hubble rate as in (\ref{IV1Bnoj}), and explicitly verify our argument that the two scalar fields scalar-tensor theory guarantees the stability of the cosmological solutions. In order to see this, we shall rewrite the FRW equations as a dynamical system, and therefore we introduce the quantities $X_{\phi}$, $X_{\chi}$ and $\bar{Y}$, which are defined in terms of the scalar fields $\phi$ and $\chi$, in the following way, 
\begin{equation}\label{newvariables}
X_{\phi}=\dot{\phi}\, , \quad  X_{\chi}=\dot{\chi}\, , \quad  \bar{Y}=\frac{\tilde{f}(\phi ,\chi)}{H}\, ,
\end{equation}
Using these, the FRW equations of Eq. (\ref{A2}) in conjunction with Eqs.~(\ref{A9}), can be written in terms of the dynamical variables (\ref{newvariables}) in the following way,
\begin{align}\label{neweqs}
& \frac{\mathrm{d}X_{\phi}}{\mathrm{d}N}=\frac{\omega '(\phi)\left (X_{\phi}^2-1\right )}{2\omega(\phi)H}-3\left (X_{\phi}-\bar{Y}\right )\nn & \frac{\mathrm{d}X_{\chi}}{\mathrm{d}N}=\frac{\eta '(\chi)\left (X_{\chi}^2-1\right )}{2\eta(\chi)H}-3\left (X_{\chi}-\bar{Y}\right ) 
\nn &\frac{\mathrm{d}\bar{Y}}{\mathrm{d}N}=\frac{3 X_{\phi} X_{\chi}\left (1-\bar{Y}^2\right )}{X_{\phi}+X_{\chi}}+\frac{\dot{H}}{H^2}\frac{X_{\phi}X_{\chi}+1-\bar{Y}(X_{\phi}+X_{\chi})}{X_{\phi}+X_{\chi}} \, .
\end{align} 
The reconstruction method solution of Eq.~(\ref{A4}), corresponds to the following values of the dynamical variables $X_{\phi}$, $X_{\chi}$ and $\bar{Y}$,
\begin{equation}\label{valuesofvariables}
X_{\phi}=1\, , \quad X_{\chi}=1\, , \quad \bar{Y}=1\, ,
\end{equation}
Thereby, in order to investigate the stability of the dynamical system of FRW equations (\ref{neweqs}), we perform linear perturbations of the new defined dynamical variables, of the following form,
\begin{equation}\label{dynamicalpertyrbationsii}
X_{\phi}=1+\delta X_{\phi} \, , \quad X_{\chi}=1+\delta X_{\chi}\, , \quad \bar{Y}=1+\delta \bar{Y}\, ,
\end{equation}
Consequently, the linear perturbations (\ref{dynamicalpertyrbationsii}) define the following dynamical system,
\begin{equation}\label{dynamicalsystemiii}
\frac{\mathrm{d}}{\mathrm{d}N}\left(
\begin{array}{c}
  \delta X_{\phi} \\
  \delta X_{\chi} \\
  \delta \bar{Y}\\
\end{array}
\right)=\left(
\begin{array}{ccc}
 -\frac{\omega' (\phi)}{H\omega (\phi)}-3 & 0 & 3
 \\ 0 &-\frac{\eta' (chi)}{H\eta (\chi)} & 3 \\
 0 & 0 & -3-\frac{\dot{H}}{H^2} \\
\end{array} \right)\left(
\begin{array}{c}
   \delta X_{\phi} \\
  \delta X_{\chi} \\
  \delta \bar{Y}\\
\end{array}
\right) \, .
\end{equation}
The matrix appearing in the dynamical system above (\ref{dynamicalsystemiii}), has the following eigenvalues,
\begin{equation}\label{eigenvaluesdoblescal}
M_{\phi}=-\frac{\omega' (\phi)}{H\omega (\phi)}-3\, , \quad {\,}M_{\chi}=-\frac{\eta' (\chi)}{H\eta (\chi)}-3\, , \quad M_{\bar{Y}}=-3-\frac{\dot{H}}{H^2} \, .
\end{equation}
Using the form of the Hubble rate given in Eq. (\ref{IV1Bnoj}), the eigenvalues become equal to,
\begin{align}\label{eigendoublehill}
& M_{\phi}=-3-\frac{2 c_3}{\left(c_4+\frac{c_1}{c_2+c_3 t}+c_5 (-t+t_s)^{\alpha }\right) (c_2+c_3 \phi )} \, ,\nn &
M_{\chi}=-3-\frac{-1+\alpha }{\left(c_4+\frac{c_1}{c_2+c_3 t}+c_5 (-t+t_s)^{\alpha }\right) (t_s-\chi )} \, ,\nn &
M_{\bar{Y}}=-3+\frac{\frac{c_1 c_3}{(c_2+c_3 t)^2}+c_5 (-t+t_s)^{-1+\alpha } \alpha }{\left(c_4+\frac{c_1}{c_2+c_3 t}+c_5 (-t+t_s)^{\alpha }\right)^2} \, ,
\end{align}
It is clear that $M_{\phi}$ and $M_{\bar{Y}}$ are negative for all $t$,  at least when the parameters $c_i$, $i=1,..5$ are chosen as in Eq. (\ref{constrmodel1}), but the eigenvalue $M_{\chi}$, can be positive when $\chi$ crosses the value $t_s$. Then it is possible that the system develops a saddle fixed point, which is a sort of mixed stability. This sort of instability could potentially indicate the presence of a new kind of physical phenomena, as we briefly discuss in a section later on.

In conclusion, the solution (\ref{A4}), with a Hubble rate as in (\ref{IV1Bnoj}), is stable, for small $\chi$ values, so during inflation. Note however that the solution (\ref{A4}) is one particular solution of the system of equations (\ref{A2}) and (\ref{A9}). In general, there might exist alternative solutions to this dynamical system, but the fact that the solution (\ref{A4}) is stable makes this solution an attractor of the dynamical system. Therefore, any class of solutions of (\ref{A2}) and (\ref{A9}), will asymptotically coincide with (\ref{A4}).

\section{$R+R^p$ limiting singular deformations of $R+aR^2$ gravity}

Having found a theoretical framework in order to describe the singular deformation of the $R^2$ gravity in the Einstein frame, in this section we shall discuss another interesting pure $F(R)$ gravity, with Jordan frame action,
\begin{equation}\label{action}
\mathcal{S}=\frac{1}{2\kappa^2}\int \mathrm{d}^4x\sqrt{-g}\left (R+\lambda R^p\right),
\end{equation}
where we ignored the contribution of all matter fluids and the parameter $p$ is not necessarily an integer. By conformally transforming the Jordan frame action, using the method we presented in the previous section, the Einstein frame canonical scalar theory is,
\be
\label{rpmodified}
S\simeq \int d^4 x \sqrt{-g}\left\{\frac{1}{2\kappa^2}R
 - \frac{1}{2}\partial_\mu \varphi \partial^\mu \varphi - \tilde{V}(\varphi)\right\}\, .
\ee
The potential $V(\varphi )$ appearing in the above equation is equal to,
\begin{equation}\label{newpotential}
V(\varphi )=V_0 e^{-2\sqrt{\frac{2}{3}}\kappa \varphi}\left(e^{\sqrt{\frac{2}{3}}\kappa \varphi}-1\right)^{\frac{p}{p-1}}
\end{equation}
with $V_0$ being equal to,
\begin{equation}\label{vo}
V_0=\frac{\kappa^2}{2}(p-1)p^{p/(1-p)}\lambda^{1/(1-p)}
\end{equation}
For an account on this model, see also \cite{sergeistarobinsky,motoyashi}. The model described by the potential (\ref{newpotential}) is a deformation of the $R^2$ inflation potential, which becomes exactly the $R^2$ inflation potential for $p=2$. In this section we shall investigate certain limits of the potential (\ref{newpotential}) and examine how a singular evolution can be linked to the resulting scalar theory, with special emphasis on the Type IV singularity. The limiting case we shall be interested in is the small $\varphi$ limit of the potential $V(\varphi)$, which is approximately equal to,
\begin{equation}\label{limitingvphi}
V(\varphi )\simeq (\kappa  \varphi )^{\frac{p}{-1+p}} \left(\left(\frac{3}{8}\right)^{\frac{p}{2-2 p}} V_0-\frac{2^{2+\frac{3}{2 (-1+p)}} 3^{-1+\frac{1}{2-2 p}} (-2+p) V_0 \kappa  \varphi }{-1+p}+\frac{2^{\frac{3 p}{2 (-1+p)}} 3^{-2+\frac{p}{2-2 p}} (12+p (-13+4 p)) V_0 \kappa ^2 \varphi ^2}{(-1+p)^2}\right)
\end{equation}
We have to note that the small $\varphi $ limit is very much related to the Type IV singularity, since we are interested in cosmic times near the Type IV singularity. As we shall see, the approximation $\varphi \rightarrow 0$, near the Type IV singularity is valid. We shall assume that $p>2$ a choice that actually corresponds to a Type IV singular evolution, as we demonstrate shortly. For $p>2$, and since we are interested in small $\varphi$ values, the most dominant term in Eq. (\ref{limitingvphi}), is the first one, therefore the scalar potential can be approximated by,
\begin{equation}\label{approxpotent}
V(\varphi )\simeq (\kappa )^{\frac{p}{-1+p}} \left(\frac{3}{8}\right)^{\frac{p}{2-2 p}} V_0 \varphi^{\frac{p}{-1+p}}
\end{equation}
As we now explicitly demonstrate, a singular evolution with a finite-time Type IV singularity may be associated to the canonical scalar theory with potential given by Eq. (\ref{approxpotent}). Indeed, consider a Hubble rate of the form,
\begin{equation}\label{hubratenew}
H(t)=f_0\left(t_s-t\right)^{\alpha}
\end{equation}
with $\alpha$ any non-integer positive number. Then, as we also discussed in a previous section, when $\alpha>1$, this corresponds to a Type IV singularity, which occurs at $t=t_s$. We shall apply the reconstruction method \cite{Nojiri:2005pu,Capozziello:2005tf} we used in section II, for the case of a single scalar field. Let the scalar-tensor theory be described by a non-canonical scalar field $\phi$, with kinetic term $\omega (\phi)$ and potential $V(\phi )$ given in Eq. (\ref{ma7einfram12e}). The by using the reconstruction method of section II, the potential reads,
\begin{equation}\label{newpotapproxnew}
V(\phi) = \frac{1}{\kappa^2}\left\{
3\left(  f_0 \left( t_s - \phi \right)^\alpha \right)^{2 \alpha }
 - \alpha f_0 \left( t_s - \phi \right)^{\alpha - 1} \right\}\, ,
\end{equation}
Since we are interested for $t\simeq t_s$, the potential (\ref{newpotapproxnew}) is approximately equal to,
\begin{equation}\label{newpotentinsert2}
V(\phi) =  - \frac{1}{\kappa^2}\left\{
 \alpha f_0 \left( t_s - \phi \right)^{\alpha - 1} \right\}\, ,
\end{equation}
In addition, the kinetic function $\omega (\phi)$ corresponding to the Hubble rate (\ref{hubratenew}) is equal to,
\begin{equation}\label{kineticfnctoin}
\omega (\phi) =- \frac{2}{\kappa^2}
\alpha f_0 \left( t_s - \phi \right)^{\alpha - 1} 
\end{equation}
By transforming the non-canonical scalar to it's canonical counterpart by using Eq. (\ref{ma13}), we obtain the following equation which relates the canonical scalar field $\varphi $, with $\phi$,
\begin{equation}\label{tsimpa}
\varphi = - \frac{2 \sqrt{2\alpha f_0}}{\kappa \left(\alpha+ 1 \right) }
\left( t_s - \phi \right)^{\frac{\alpha + 1}{2}}
\end{equation}
This relation validates our small $\varphi $ approximation, since as the cosmic time approaches the singularity, that is $t\rightarrow t_s$, then the canonical scalar tends to zero. By substituting this to the scalar potential $V(\phi)$ given in Eq. (\ref{newpotapproxnew}), we obtain the scalar potential in terms of the canonical scalar field,
\begin{equation}\label{neweqnarhcdld}
V(\phi (\varphi ))\simeq - \frac{\alpha f_0 }{\kappa^2}
\left\{ - \frac{\kappa \left(\alpha+ 1 \right) }{2 \sqrt{2\alpha f_0}} 
\right\}^{\frac{2 \left( \alpha - 1 \right)}{\alpha + 1}}\varphi ^{\frac{2 \left( \alpha - 1 \right)}{\alpha + 1}}
\end{equation}
Thus by looking Eqs. (\ref{neweqnarhcdld}) and (\ref{newpotential}), the potentials become identical if we make the following identifications,
\begin{equation}\label{leftidentifications}
(\kappa )^{\frac{p}{-1+p}} \left(\frac{3}{8}\right)^{\frac{p}{2-2 p}} V_0 =  \frac{\alpha f_0 }{\kappa^2}
\left\{  \frac{\kappa \left(\alpha+ 1 \right) }{2 \sqrt{2\alpha f_0}} 
\right\}^{\frac{2 \left( \alpha - 1 \right)}{\alpha + 1}},{\,}{\,}{\,}{\,}{\,}\frac{p}{p-1}=\frac{2 \left( \alpha - 1 \right)}{\alpha + 1}
\end{equation}
Thus by the second relation of Eq. (\ref{leftidentifications}), we may conclude that when $p>2$, a Type IV singularity occurs. Therefore, we successfully related a Type IV singular evolution to the scalar theory with canonical scalar potential given in (\ref{newpotential}). What remains is to examine the stability of the solution (\ref{ma11}), which in our case is,
\begin{equation}\label{newsolcheck}
\phi=t\, ,\quad H=f(t)=f_0\left (t_s-t\right)^{\alpha }\, .
\end{equation}
The stability of this solution is the subject of the next section.

\subsubsection{Stability analysis of single scalar field case}

As we pointed out in a previous section, when the single scalar field reconstruction method is employed, infinite instabilities frequently occur at the time instance that the phantom divide $w_\mathrm{eff}=-1$ is crossed \cite{Ito:2011ae,vikman}. We shall examine whether instabilities occur in the case of the solution given in Eq. (\ref{newsolcheck}). In order to study the stability of the solution, it is convenient to rewrite the FRW equations in terms of a dynamical system. To this end, we introduce the variables, $X_{\phi}$ and $Y$, which are defined as follows,
\begin{equation}
\label{newvaribles}
X_{\phi}=\dot{\phi}\, , \quad Y=\frac{f(\phi)}{H}\, .
\end{equation}
Notice that, from the form of the variable $Y$, we can realize that practically it measures the deviation from the reconstruction solution (\ref{ma11}). Making use of these variables and upon combining the FRW equation and the field equation for the scalar field $\phi$, we obtain the following dynamical system,
\be
\frac{\mathrm{d}X_{\phi}}{\mathrm{d}N}=\frac{f''(\phi)\left (X_{\phi}^2-1\right 
)}{2f'(\phi)H}-3\left (X_{\phi}-Y\right )\, , \quad 
\frac{\mathrm{d}Y}{\mathrm{d}N}=\frac{f'(\phi)\left(1-X_{\phi}Y\right)X_{\phi}}{H^2}\, ,
\label{newfrweqns}
\ee
where $N$ stands for the $e$-folding number. We may easily verify that the solution of 
Eq.~(\ref{ma11}), has a direct correspondence to the following values of the variables $X_{\phi}$ and $Y$,
\begin{equation}
\label{standardvalues}
X_{\phi}=1\, ,\quad Y=1\, ,
\end{equation}
So practically the point $(1,1)$ in the $(X,Y)$ plane is the critical point of the dynamical system (\ref{newfrweqns}). The stability of the dynamical system may be revealed if the system is perturbed linearly around the critical point $(1,1)$,
\begin{equation}
\label{linearpert}
X_{\phi}=1+\delta X_{\phi}\, , \quad Y=1+\delta Y\, ,
\end{equation}
and by doing so, the dynamical system (\ref{newfrweqns}) can be cast as follows,
\begin{equation}
\label{dynamicalsystem}
\frac{\mathrm{d}}{\mathrm{d}N}\left(
\begin{array}{c}
  \delta X_{\phi} \\
  \delta Y \\
\end{array}
\right)=\left(
\begin{array}{cc}
 -\frac{\ddot{H}}{\dot{H}H}-3 & 3
 \\ -\frac{\dot{H}}{H^2} & -\frac{\dot{H}}{H^2} \\
\end{array} \right)\left(
\begin{array}{c}
  \delta X_{\phi} \\
  \delta Y \\
\end{array}
\right)\, .
\end{equation}
It is a textbook known fact in dynamical systems theory \cite{jost}, that the dynamical system (\ref{dynamicalsystem}) can be considered stable, if the eigenvalues of the matrix $M$,
\begin{equation}\label{matrixm}
M=\left(
\begin{array}{cc}
 -\frac{\ddot{H}}{\dot{H}H}-3 & 3
 \\ -\frac{\dot{H}}{H^2} & -\frac{\dot{H}}{H^2} \\
\end{array} \right)
\end{equation}
are negative. A simple analytic calculation of the eigenvalues yields, 
\begin{align}
\label{eigenvalues}
M_+=&\frac{1}{2} \left[-\left(\frac{H''(t)}{H'(t) 
H(t)}+\frac{H'(t)}{H(t)^2}+3\right) \right. \nn 
& \left. +\sqrt{\left(\frac{H''(t)}{H'(t) 
H(t)}+\frac{H'(t)}{H(t)^2}+3\right)^2-\frac{4 H''(t)}{H(t)^3}-\frac{12 H'(t)}{H(
t)^2}}\right] \, , \nn
M_{-}=&\frac{1}{2} \left[-\left(\frac{H''(t)}{H'(t) 
H(t)}+\frac{H'(t)}{H(t)^2}+3\right) \right. \nn 
& \left. -\sqrt{\left(\frac{H''(t)}{H'(t) 
H(t)}+\frac{H'(t)}{H(t)^2}+3\right)^2-\frac{4 H''(t)}{H(t)^3}-\frac{12 H'(t)}{H(
t)^2}}\right] \, ,
\end{align}
In the case of the Hubble rate given in Eq. (\ref{newsolcheck}), the eigenvalues become in general,
\begin{align}
\label{eigenvalues1}
& M_+= \frac{1}{2}\Big{(} -3+\frac{(-t+t_s)^{-1-\alpha } (-1+\alpha )}{f_0}+\frac{(-t+t_s)^{-1-\alpha } \alpha }{f_0}      \Big{)}  
\\ \notag & + \frac{1}{2}\Big{(} \sqrt{ \left(\frac{12 (-t+t_s)^{-1-\alpha } \alpha }{f_0}-\frac{4 (-t+t_s)^{-2-2 \alpha } (-1+\alpha ) \alpha }{f_0^2}+\left(3-\frac{(-t+t_s)^{-1-\alpha } (-1+\alpha )}{f_0}-\frac{(-t+t_s)^{-1-\alpha } \alpha }{f_0}\right)^2\right)}\Big{)}\\ \notag &
M_{-}= \frac{1}{2} \left(-3+\frac{(-t+t_s)^{-1-\alpha } (-1+\alpha )}{f_0}+\frac{(-t+t_s)^{-1-\alpha } \alpha }{f_0}\right.
\\ \notag & -\sqrt{\left(\frac{12 (-t+t_s)^{-1-\alpha } \alpha }{f_0}-\frac{4 (-t+t_s)^{-2-2 \alpha } (-1+\alpha ) \alpha }{f_0^2}+\left(3-\frac{(-t+t_s)^{-1-\alpha } (-1+\alpha )}{f_0}-\frac{(-t+t_s)^{-1-\alpha } \alpha }{f_0}\right)^2\right)}
\end{align}
Although by appropriately choosing $\alpha$, and also by assuming that $t$ is away from the singular point, the above eigenvalues can be negative. But as the cosmic time approaches $t_s$, the eigenvalues have an infinite instability, with the eigenvalues being positive in some cases. This makes the system unstable at the transition point. Note however that instabilities may serve as indicators that the evolutionary process may change at the time these occur, as we briefly discuss in the next section. 

Before we close this section, for completeness we shall briefly provide a description with two scalar fields, one of which is a non-canonical field. We presented the essentials of the two scalar field method in a previous section, so the action is given by Eq. (\ref{A1}), the kinetic functions by (\ref{A5}), while the scalar potential is given by Eq. (\ref{A8}). By choosing the auxiliary function $\alpha(x)$ as follows,
\begin{equation}\label{ax}
\alpha(x)=f_0 (-x +t_s)^{-1+\alpha } \alpha
\end{equation}
we get the kinetic functions of the scalar field $\phi$ to be,
\begin{equation}\label{omegafifun}
\omega (\phi)=\frac{2 f_0 \alpha  (t_s-\phi )^{-1+ \alpha }\left(1+\sqrt{2}\right)}{\kappa ^2}
\end{equation}
while the kinetic function of the scalar field $\chi$ reads,
\begin{equation}\label{omegafifun1}
\eta (\chi)= -\frac{2 \sqrt{2} f_0 \alpha  (t_s-\chi )^{-1+ \alpha }}{\kappa ^2}
\end{equation}
The corresponding scalar potential $V(\phi,\chi )$ is given by,
\begin{align}\label{scalarpotneone}
& V(\phi,\chi )=\frac{\sqrt{b_0 \alpha } (t_s-\phi )^{\frac{1}{2} (-1+\alpha )}+b_0 \alpha  (t_s-\phi )^{-1+\alpha }+\sqrt{b_0 \alpha } (t_s-\chi )^{\frac{1}{2} (-1+\alpha )}}{\kappa ^2} \\ \notag & 
+\frac{3 \left(b_0 (1+\alpha ) (t_s-\phi )^{\alpha }+2 \sqrt{b_0 \alpha } \left((t_s-\phi )^{\frac{1+\alpha }{2}}+(t_s-\chi )^{\frac{1+\alpha }{2}}\right)\right)^2}{(1+\alpha )^2 \kappa ^2}
\end{align}

\section{Inclusion of Matter Fluids in the Cosmological Evolution}

Having studied the cosmological evolution of Eq. (\ref{IV1Bnoj}) in the absence of matter fluids, in this section we shall include the contribution of perfect matter fluids. Particularly, we shall include the effects of pressure-less matter, but in principle other matter fluids can be considered too. We start off with the description of the equations of motion in the presence of matter fluids, with constant equation of state $w_m$, which take the following form,
\begin{align}\label{gfgdjsk}
& H^2=\frac{\kappa^2}{3}\left( \rho_m+\frac{1}{2}\omega (\phi)\dot{\phi}^2+\frac{1}{2}\eta (\chi)\dot{\chi}^2+V(\phi, \chi )\right),\\ \notag &
\dot{H}=-\frac{\kappa^2}{2}\left( \rho_m+p_m+\omega (\phi)\dot{\phi}^2+\eta (\phi)\dot{\chi}^2\right)\, ,
\end{align}
where $\rho_m$ and $p_m$ stand for the effective energy density and effective pressure of the matter fluids with equation of state $p_m=w_m\rho_m$. The effective energy density and pressure satisfy the conservation law,
\begin{equation}\label{conserv}
\dot{\rho_m}+3 H\left (\rho_m+p_m\right )\, ,
\end{equation}
from which we get $\rho_m=\rho_{m_0}a^{-3 (1+w_m)}$. Now consider the following solution to the FRW equations (\ref{gfgdjsk}),
\begin{equation}\label{gfkska}
\phi=\chi=t,\,\,\, H=f(t)\, .
\end{equation} 
 Then then kinetic terms $\omega (\phi)$ and $\eta (\chi )$, satisfy,
 \begin{equation}\label{omegaandchi}
 \omega (t)+\chi (t)=-\frac{2}{\kappa^2}f'(t)+\left(w_m+1\right)F_0e^{\left(1+w_m\right)F(t)}\, ,
\end{equation}
where $F(t)=f'(t)$. The last term follows from the fact that $\rho_m=\rho_{m_0}a^{-3 (1+w_m)}$, and also by combining the first relation of Eq. (\ref{gfgdjsk}) and $H=\dot{a}/a$. So in general, the kinetic terms that generate the cosmological evolution (\ref{gfkska}), can be chosen as,
\begin{align}\label{twosol}
& \omega (\phi)= -\frac{2}{\kappa^2}\left(f'(\phi)-\sqrt{a_1(\phi)^2+f'(\phi)^2}\right)+\frac{\left(w_m+1\right)}{2}F_0e^{\left(1+w_m\right)F(\phi)}, \\ \notag & \eta (\chi)=-\frac{2}{\kappa^2}\left(\sqrt{a_1(\phi)^2+f'(\phi)^2}\right)+\frac{\left(w_m+1\right)}{2}F_0e^{\left(1+w_m\right)F(\chi)}\, .
\end{align}
where $a_1(x)$ is an arbitrary function. Indeed, it can be easily verified that the above two kinetic terms satisfy Eq. (\ref{omegaandchi}), when $\phi=\chi=t$. Also we define the function,
\begin{equation}\label{tildefwithmat}
\tilde{f}(\phi,\chi)=-\frac{\kappa^2}{2}\left(\int \omega (\phi) \mathrm{d}\phi +\int \eta (\chi) \mathrm{d}\chi-\int f_1(\phi)\mathrm{d}\phi-\int f_1(\chi)\mathrm{d}\chi\right)\, ,
\end{equation}
with $f_1(x)=\frac{\left(w_m+1\right)}{2}F_0e^{\left(1+w_m\right)F(x)}$. The function $\tilde{f}(\phi,\chi)$ satisfies $\tilde{f}(t,t)=f(t)$, as it can be easily verified since,
\begin{equation}\label{satisfsecoeqn}
\int \left(\omega (t)+\eta (t)-f_1(t)-f_1(t) \right )\mathrm{d}t=\int f'(t)\mathrm{d}t=f(t)\, ,
\end{equation}
which holds true, due to Eq. (\ref{omegaandchi}). So the potential that generates the cosmological evolution (\ref{gfkska}), can be chosen as follows,
\begin{align}\label{potentialmatter}
& V(\phi,\chi)=\frac{1}{\kappa^2}\left(3\tilde{f}(\chi,\phi)^2+\frac{\partial \tilde{f}}{\partial \phi}+\frac{\partial \tilde{f}}{\partial \chi}\right) -\frac{\left(w_m-1\right)}{4}F_0e^{-3(1+w_m)F(\phi)}\\ \notag & -\frac{\left(w_m-1\right)}{4}F_0e^{-3(1+w_m)F(\chi)}\, .
\end{align}
We chose the potential as in Eq. (\ref{potentialmatter}), so that at $\phi=\chi=t$ it satisfies,
\begin{equation}\label{potentsatisfieds}
V(t,t)=\frac{1}{\kappa^2}\left( f(t)^2+f'(t)\right)-\frac{\left(w_m-1\right)}{2}F_0e^{-3(1+w_m)F(t)}\, .
\end{equation}
It is easy to prove that the potential $V(\phi,\chi)$ of Eq. (\ref{potentialmatter}), since at $\phi=\chi=t$, the following relations hold true:
\begin{align}\label{aligeqnsa}
& 3 \tilde{f}(t,t)=3f(t)^2,\\ \notag & \frac{\partial \tilde{f}}{\partial \phi}\Big{|}_{\phi=t}=\omega (t)-f_1(t),\\ \notag &
\frac{\partial \tilde{f}}{\partial \chi}\Big{|}_{\chi=t}=\chi (t)-f_1(t)\, 
\end{align}
and consequently at $\chi=\phi=t$, we have,
\begin{equation}\label{cft}
\frac{\partial \tilde{f}}{\partial \phi}\Big{|}_{\phi=t}+\frac{\partial \tilde{f}}{\partial \chi}\Big{|}_{\chi=t}=\eta (t)+\omega (t)-2f_1(t)=-\frac{2}{\kappa^2}f'(t)\, ,
\end{equation}
where we also used Eq. (\ref{omegaandchi}). Let us find the corresponding kinetic terms $\omega (\phi)$, $\eta (\chi)$, the function $\tilde{f}$ and the potentia $V(\phi,\chi)$, for the cosmological evolution of Eq. (\ref{IV1Bnoj}). By combining Eqs. (\ref{IV1Bnoj}) and (\ref{twosol}), we get the kinetic terms $\omega (\phi)$ and $\eta (\chi)$, which are,
\begin{align}\label{frix}
& \omega (\phi)=\frac{1}{2} e^{-3 (1+w_m) \left(-\frac{c_5 (t_s-\phi )^{1+\alpha }}{1+\alpha }+c_4 \phi +\frac{c_1 }{c_3}\ln (c_2+c_3 \phi )\right)} F_0 (1+w_m)+\frac{2 c_1 c_3}{\kappa ^2 (c_2+c_3 \phi )^2} \\ \notag &
\eta (\chi)= \frac{1}{2} e^{-3 (1+w_m) \left(-\frac{c_5 (t_s-\chi )^{1+\alpha }}{1+\alpha }+c_4 \chi +\frac{c_1 }{c_3}\ln (c_2+c_3 \chi )\right)} F_0 (1+w_m)-\frac{2 c_2 \alpha  (t_s-\chi )^{-1+\alpha }}{\kappa ^2}\, ,
\end{align}
while the potential of Eq. (\ref{potentialmatter}), becomes equal to,
\begin{align}\label{dfere}
& V(\phi,\chi)=-\frac{1}{4} e^{-3 (1+w_m) \left(c_4 \phi +c_5 (t_s-\phi )^{\alpha } \left(-\frac{t_s}{1+\alpha }+\frac{\phi }{1+\alpha }\right)+\frac{c_1 }{c_3}\ln (c_2+c_3 \phi )\right)} F_0\\ \notag & -\frac{1}{4} e^{-3 (1+w_m) \left(c_4 \chi +c_5 (t_s-\chi )^{\alpha } \left(-\frac{t_s}{1+\alpha }+\frac{\chi }{1+\alpha }\right)+\frac{c_1 }{c_3}\ln (c_2+c_3 \chi )\right)} F_0\\ \notag &
\frac{1}{4} e^{-3 (1+w_m) \left(c_4 \phi +c_5 (t_s-\phi )^{\alpha } \left(-\frac{t_s}{1+\alpha } +\frac{\phi }{1+\alpha }\right)+\frac{c_1 }{c_3}\ln (c_2+c_3 \phi )\right)} F_0 w_m\\ \notag &+\frac{1}{4} e^{-3 (1+w_m) \left(c_4 \chi +c_5 (t_s-\chi )^{\alpha } \left(-\frac{t_s}{1+\alpha }+\frac{\chi }{1+\alpha }\right)+\frac{c_1 }{c_3}\ln (c_2+c_3 \chi )\right)} F_0 w_m\\ \notag &
+\frac{3 c_4^2}{\kappa ^2}+\frac{3 c_1^2}{\kappa ^2 (c_2+c_3 \phi )^2}-\frac{c_1 c_3}{\kappa ^2 (c_2+c_3 \phi )^2}+\frac{6 c_1 c_4}{\kappa ^2 (c_2+c_3 \phi )}\\ \notag &
-\frac{c_5 \alpha  (t_s-\chi )^{-1+\alpha }}{\kappa ^2}+\frac{6 c_4 c_5 (t_s-\chi )^{\alpha }}{\kappa ^2}+\frac{6 c_1 c_5 (t_s-\chi )^{\alpha }}{\kappa ^2 (c_2+c_3 \phi )}+\frac{3 c_5^2 (t_s-\chi )^{2 \alpha }}{\kappa ^2}\, .
\end{align}
We need to note we chose the function $a_1(x)$ as in Eq. (\ref{alphaxnewdef}). For an application of the method we used to include matter fluids, but in the case of a single scalar field, the reader is referred to Ref. \cite{sergnoj}.

\section{Significance of the Type IV Singularity}

In the previous sections we investigated how a Type IV singularity can be connected to the cosmological evolution of the $R^2$ inflation model. As we discussed this can be done only with the use of a second scalar field, which is dynamically insignificant at early-time. However, we did not discuss our motivation for studying such a singular evolution. In principle it could be claimed that a Type IV singularity is not a real singularity, but now we provide some arguments to support our study and the significance of the Type IV singularity. Firstly, it is not a crushing type singularity so the Universe can smoothly pass such a singular point and continue undisturbed its evolution. Secondly and more importantly, the Type IV singularity can affect the observational indices of inflation in a dramatic way, if the slow-roll condition is abandoned. As we will demonstrate, the effect of a Type IV singularity is to cause an instability to the dynamical system, which can indicate that the attractor solution that drives the early-time evolution is unstable, and in turn this could be an indicator that inflation ends and graceful exit is achieved. In the rest of this section we thoroughly discuss this issue, but for a similar situation consult Ref. \cite{oikonomouodigrexit}.

We shall demonstrate that the Hubble second slow-roll parameter $\eta_H$ develops an instability at the point where a Type IV singularity occurs. Recall that the second Hubble slow-roll parameter is equal to \cite{barrowslowroll}:
\begin{equation}\label{secondhiusbbleslow}
\eta_H=-\frac{\ddot{H}}{2H\dot{H}}\, ,
\end{equation}
and in inflationary dynamics, the second Hubble slow-roll parameter measures how much does inflation lasts, while the first Hubble slow-roll index measures if inflation occurs in the first place. Note that we abandon the slow-roll condition for all the scalar fields of our model. Let us calculate the second Hubble slow-roll index for the Hubble rate appearing in Eq. (\ref{IV1Bnoj}), assuming that $\alpha=4/3$ or more generally that $1<\alpha <2$, as in the previous sections. For the Hubble rate of Eq. (\ref{IV1Bnoj}), the second Hubble rate (\ref{secondhiusbbleslow}) becomes,

\begin{equation}\label{etahub}
\eta_H=-\frac{\frac{2 c_1 c_3^2}{(c_2+c_3 t)^3}+c_5 (-t+t_s)^{-2+\alpha } (-1+\alpha ) \alpha }{2 \left(c_4+\frac{c_1}{c_2+c_3 t}+c_5 (-t+t_s)^{\alpha }\right) \left(-\frac{c_1 c_3}{(c_2+c_3 t)^2}-c_5 (-t+t_s)^{-1+\alpha } \alpha \right)}{\,}.
\end{equation}
By looking Eq. (\ref{etahub}), we can easily observe that the term $\sim (-t+t_s)^{-2+\alpha }$ appearing in the numerator of the fraction, becomes divergent at $t=t_s$, for the values of $\alpha$ we assumed. This clearly indicates an infinite instability of the inflationary dynamics. In order to make things more clear, let us recall the significance of the inflationary indices and their interpretation. We follow the analysis of Ref. \cite{barrowslowroll}. As we demonstrate, this instability could be viewed as an obstruction in the inflationary evolution and hence can act as a mechanism for graceful exit from inflation, or at lest as an indicator of exit from inflation. However, this mechanism is different in spirit from other mechanism for graceful exit, nevertheless the instability occurs, so we should analyze it's significance.

A clear interpretation of the slow-roll indices was provide in Ref. \cite{barrowslowroll}, where the slow-roll expansion was used. The inflationary dynamics is determined by the slow-roll expansion, and particularly the slow-roll expansion indicates if an inflationary attractor is the correct inflationary solution. It is a perturbation expansion, the first terms of which are the usual slow-roll indices. Note that the slow-roll expansion is a more strong and restrictive physical description of inflation, in comparison to the slow-roll approximation \cite{barrowslowroll}. Let us briefly describe the slow-roll expansion by using a general example. Consider an inflationary solution described by a Hubble evolution $H(\varphi )$, with $\varphi$ being a canonical scalar field. The Hubble slow-roll expansion determines an inflationary solution which is an asymptotic attractor of all the inflationary solutions obtained by the usual potential slow-roll approximation. The FRW can be written as,
\begin{equation}\label{frqwe}
H^2(\varphi )=\frac{8 \pi\kappa^2}{3}V(\varphi )\Big{(}1-\frac{1}{3}\epsilon_H (\varphi) \Big{)}^{-1}\, ,
\end{equation} 
where $\epsilon_H$ is the first Hubble slow-roll parameter,
\begin{equation}\label{fhubslowroll}
\epsilon_H=-\frac{\dot{H}}{H^2}\, ,
\end{equation}
but in Eq. (\ref{frqwe}) is expressed as a function of the scalar field $\varphi$. The Hubble slow-roll expansion is obtained if the FRW equation is expanded in a perturbation series, 
\begin{align}\label{frqwe1}
& H^2(\varphi )\simeq \frac{8 \pi\kappa^2}{3}V(\varphi )\Big{(} 1+\epsilon_V-\frac{4}{3}\epsilon_V^2+\frac{2}{3}\epsilon_V\eta_V
\\ & \notag +\frac{32}{9}\epsilon_V^3+\frac{5}{9}\epsilon_V\eta_V^2-\frac{10}{3}\epsilon_V^2\eta_V+\frac{2}{9}\epsilon_V\xi^2_V+\mathcal{O}_4
 \Big{)}\, ,
\end{align} 
where terms up to fourth order are kept, and the parameters $\epsilon_V,\eta_V,\xi_V$ are given in terms of the Hubble slow-roll parameters $\eta_H$ and $\epsilon_H$ as follows,
\begin{align}\label{shelloween}
& \eta_V=(3-\epsilon_H)^{-1}\Big{(}3\epsilon_H+3\eta_H-\eta_H^2-\xi_H^2\Big{)} ,
\\ & \notag \xi_V=(3-\epsilon_H)^{-1}\Big{(}27\epsilon_H\eta_H+9\xi_H^2-9\epsilon_H\eta_H^2-12\eta_H\xi_H^2-3\sigma_H^3+3\eta_H^2\xi_H^2+\eta_H\sigma_H^3 \Big{)}\, .
\end{align}
In addition, $\xi_H$ and $\sigma_H$ are defined in terms of $\epsilon_H$ and $\eta_H$ below,
\begin{equation}\label{insertcode123sledgehammerenteraccept}
\xi_H^2=\epsilon_H\eta_H-\sqrt{\frac{1}{4\pi\kappa^2}}\sqrt{\epsilon_H}\eta_H',\,\,\,\sigma_H^3=\xi_H^2(2\epsilon_H-\eta_H)-\sqrt{\frac{1}{\kappa^2\pi}}\sqrt{\epsilon_H}\xi_H\xi_H'\, .
\end{equation}
In the above equation, the prime indicates differentiation with respect to the scalar field $\varphi$. It is conceivable that the Hubble slow-roll expansion we just described is a better approximation to the final inflationary attractor $H(t)$, in comparison to the standard slow-roll approximation, which uses only the parameters $\epsilon$ and $\eta_V$ at lowest order. Note that the potential slow-roll parameters correspond to the ones defined in Eq. (\ref{slowrolpparaenedefs}), in terms of the potential of the canonical scalar field. In addition, a useful relation that we will make use in the sections to follow, is that at lowest order, the Hubble slow-roll parameters $\epsilon_H,\eta_H$ and the potential slow-roll parameters $\epsilon_V,\eta_V$ for a canonical scalar field are related as follows, 
\begin{equation}\label{hgfgsdgs}
\epsilon_H=\epsilon_V,\,\,\, \eta_H=\eta_V-\epsilon\, .
\end{equation}
Coming back to the Hubble slow-roll expansion interpretation, as was also discussed by the authors of \cite{barrowslowroll}, when the slow-roll perturbative expansion breaks down, the inflationary solution ceases to be the final attractor of the theory, thus inflation ends. This breakdown of the perturbative expansion can occur when the slow-roll parameters take large values, or if a singularity occurs. The latter is the case for a Type IV singularity. Hence our motivation for studying the Type IV singularity is exactly this non-trivial feature of the inflationary dynamics.

\section{Observational Indices with the Double Scalar Formalism}

In this paper we made a crucial assumption, with regards to the early-time dynamics of the scalar fields $\varphi$ and $\chi$, which was that the scalar field $\chi$ does not make a significant contribution at early-time. Thus the contribution of the scalar field, in the potential and also its kinetic term, can be safely disregarded. This result was supported numerically in the previous sections, however, we need to further support this by using the multi-scalar field formalism for the calculation of the slow-roll and observational indices. We need to stress that when both scalar fields are taken into account, the space of trajectories is much more large in comparison to the single scalar field case. With the present investigation we aim to demonstrate that when the parameters are of the same order as the ones given in Eq. (\ref{parameters}), then the observational indices and slow-roll parameters lead to the same result as in the single scalar field case. We shall use two approaches with regards to the slow-roll parameters, the one developed in Ref. \cite{kaizer}, which is related to the slow-roll condition, and the second one is related to the so-called Hubble slow-roll parameters, developed in Ref. \cite{barrowslowroll}. Note that the first slow-roll parameter $\epsilon$ coincides in both the methods we shall use. However the second slow-roll parameter is different in the two approaches. For more detail consult Ref. \cite{barrowslowroll}.

We use the formalism and notation of Ref. \cite{kaizer}. We consider the multi-scalar field action,
\begin{equation}\label{miltuiscale}
S=\int \mathrm{d}^4x\sqrt{-\hat{g}}\Big{ (} \frac{\hat{R}}{2\kappa^2}-\frac{1}{2}G_{IJ}(\phi^I)\hat{g}^{\mu \nu }\partial_{\mu }\phi^I\partial_{\nu }\phi^J -V(\phi^{I} ) \Big{)}\, ,
\end{equation} 
with $I,J=1,2$. In our case the scalar fields $\phi^I$ are $\phi^1=\varphi$ and $\phi^2=\chi$. Moreover, the metric of the scalar field configuration space $G_{IJ}(\phi^I)$, depends on the scalar fields. In our case, the field space is two dimensional, and the matrix representation of the metric for the two scalar field action of Eq. (\ref{A1new}), equals to,
\begin{equation}\label{metricconfspace}
G=\left(
\begin{array}{cc}
 1 & 0 \\
 0 & \frac{2 c_2 \alpha  (\chi-t_s )^{-1+\alpha }}{\kappa ^2} \\
\end{array}
\right)\, .
\end{equation}
The FRW equations for the action (\ref{miltuiscale}) are,
\begin{align}\label{gfgdgd}
& H^2=\frac{\kappa^2}{3}\left( \frac{1}{2}G_{IJ}\dot{\varphi}^I\dot{\varphi}^J+V(\varphi^I)\right)\\ \notag &
\dot{H}=-\frac{1}{\kappa^2}G_{IJ}\dot{\varphi}^I\dot{\varphi}^J\\ \notag &
\square \phi^I+\hat{g}^{\mu \nu }\Gamma^I_{IK}\partial_{\mu }\phi^J\partial_{\nu }\phi^K-G^{IK}V_{,K}=0\, ,
\end{align}
with $V_{,K}=\partial V/\partial \phi^K$, and also with $\Gamma^I_{IK}=\Gamma^I_{IK}(\varphi,\phi)$ we denote the Christoffel symbols corresponding to the two dimensional scalar field configuration space, with the metric being the one given in Eq. (\ref{metricconfspace}). For this metric, the only non-zero Christoffel symbol is,
\begin{align}\label{christofellsymbols}
\Gamma^{\chi}_{\chi \chi }= \frac{-1+\alpha }{2 (-t_s+\chi )} \, .
\end{align}

For the sake of notational simplicity, we introduce the scalar field $\sigma$ and the vector field $\hat{\sigma}^{I}$, which are defined to be,
\begin{equation}\label{dsigmadot}
\dot{\sigma}=\sqrt{G_{IJ}\dot{\phi}^{I}\dot{\phi}^{J}},\,\,\, \hat{\sigma}^{I}=\frac{\dot{\phi}^I}{\dot{\sigma}}.
\end{equation}
In terms of $\dot{\sigma}$ and $\hat{\sigma}^{I}$, the field equations (\ref{gfgdgd}) are written as follows,
\begin{align}\label{fieldhf}
& H^2=\frac{\kappa^2}{3}\left(\frac{1}{2}\dot{\sigma}^2+V \right), \\ \notag &
\dot{H}=-\frac{\kappa^2}{2}\dot{\sigma}^2,\\ \notag &
\ddot{\sigma}+3H\dot{\sigma}+V_{,\sigma}=0
\end{align}
where $V_{,\sigma}=\hat{\sigma}^IV_{,I}$. In our case, the potential $V$ of Eqs. (\ref{gfgdgd}) and (\ref{fieldhf}) is identified with the potential of Eq. (\ref{finalpotential}). Since $\dot{\sigma}$ will be extensively used in the following, it is worth computing it, and for the metric of Eq. (\ref{metricconfspace}), it reads,
\begin{align}\label{dhfhf}
& \dot{\sigma}=\sqrt{\dot{\varphi}^2-\frac{2 c_2 \alpha  (-t_s+\chi )^{-1+\alpha }}{\kappa ^2}\dot{\chi}^2}\, .
\end{align}
Up to this point, the potential slow-roll approach with two scalars and the Hubble slow-roll approach is the same. However, we differentiate our analysis here, since the second slow-roll parameter is different when calculated in the context of the two aforementioned approaches. We start of with the potential slow-roll parameters.

\subsubsection{Potential Slow-Roll Parameters Approach}

We calculated the slow-roll indices in the potential slow-roll approach. We use this name, since as was shown in \cite{kaizer}, the indices are related to the scalar potential directly. However, we shall use the slow-roll limit for the canonical scalar field $\varphi$, and also we shall take into account the values of the parameters as in Eq. (\ref{parameters}). As we now demonstrate, the observational indices are approximately equal to the ones corresponding to the single scalar field case.

In both cases, the first slow-roll parameter $\epsilon$, is equal to,
\begin{equation}\label{firstslowroll}
\epsilon =-\frac{\dot{H}}{H^2}=\frac{3\dot{\sigma}^2}{\dot{\sigma}^2+\tilde{V}}\, ,
\end{equation}
We can easily calculate this for the potential of Eq. (\ref{finalpotential}), so the parameter $\epsilon$ reads,
\begin{equation}\label{sgdgd}
\epsilon= 3 \frac{\dot{\varphi}^2-\frac{2 c_2 \alpha  (-t_s+\chi )^{-1+\alpha }}{\kappa ^2}\dot{\chi}^2}{\dot{\varphi}^2-\frac{2 c_2 \alpha  (-t_s+\chi )^{-1+\alpha }}{\kappa ^2}\dot{\chi}^2+\tilde{V(\varphi,\chi)}}\, .
\end{equation}
For the values of the parameters appearing in Eq. (\ref{parameters}), we can simplify the parameter $\epsilon$, since the following hold true,
\begin{align}\label{simlpify}
& \dot{\varphi}^2-\frac{2 c_2 \alpha  (-t_s+\chi )^{-1+\alpha }}{\kappa ^2}\dot{\chi}^2\simeq \dot{\varphi}^2, \\ \notag &
\tilde{V(\varphi,\chi)}\simeq V(\varphi)
\end{align}
 with $V(\varphi)$ given in Eq. (\ref{vsnew}). Therefore, it simply follows that $\epsilon$ equals to,
 \begin{equation}\label{egdfbarrows}
 \epsilon =3\frac{\dot{\varphi}^2}{\dot{\varphi}^2+2V(\varphi)}=3\frac{\frac{\dot{\varphi}^2}{2}}{\frac{\dot{\varphi}^2}{2}+V(\varphi)}\, .
\end{equation}
 Recalling that we assumed that the canonical scalar field $\varphi$ satisfies the slow-roll conditions, then as was shown in \cite{barrowslowroll}, the parameter $\epsilon$ equals to,
 \begin{equation}\label{dgfaae}
 \epsilon=\epsilon_H=3\frac{\frac{\dot{\varphi}^2}{2}}{\frac{\dot{\varphi}^2}{2}+V(\varphi)}\simeq \frac{1}{2\kappa^2}\left(\frac{V '(\varphi )}{V (\varphi
)}\right)\, .
\end{equation}
 Note that this results holds true at first order, as we briefly explained in the previous section. Now we proceed to the second slow-roll parameter $\eta$. We denoted this as $\eta$ in order to discriminate this from the Hubble slow-roll parameter $\eta_H$, which we compute in the next subsection. Since the parameter $\epsilon$ takes small values, as was shown in \cite{kaizer}, the potential slow-roll index $\eta$ is approximately equal to,
\begin{equation}\label{etaslowrollahsgdg}
\eta \simeq \epsilon -\frac{\ddot{\sigma}}{H\dot{\sigma}}+\mathcal{O}(\epsilon^2)\, .
\end{equation}
We can easily calculate $\ddot{\sigma}$ from Eq. (\ref{dsigmadot}) and it reads,
\begin{equation}\label{voithitiki}
\ddot{\sigma}= \frac{\frac{\mathrm{d}G_{IJ}}{\mathrm{d}t}\dot{\phi}^I\dot{\phi}^J+G_{IJ}\ddot{\phi}^I\dot{\phi}^J+G_{IJ}\dot{\phi}^I\ddot{\phi}^J}{2\sqrt{G_{IJ}\dot{\phi}^I\dot{\phi}^J}}\, ,\, ,
\end{equation}
so in our case it is equal to,
\begin{align}\label{voithia2}
& \ddot{\sigma}=-\frac{4 c_2 (-1+\alpha ) \alpha  (-t_s+\chi )^{-2+\alpha } \dot{\chi}}{\kappa ^22\sqrt{\dot{\varphi}^2-\frac{2 c_2 \alpha  (-t_s+\chi )^{-1+\alpha }\dot{\chi}^2}{\kappa ^2}}}
+\frac{2\dot{\varphi}\ddot{\varphi}}{2\sqrt{\dot{\varphi}^2-\frac{2 c_2 \alpha  (-t_s+\chi )^{-1+\alpha }\dot{\chi}^2}{\kappa ^2}}}
 -\frac{4 c_2 \alpha  (-t_s+\chi )^{-1+\alpha }\ddot{\chi}\dot{\chi}}{\kappa ^22\sqrt{\dot{\varphi}^2-\frac{2 c_2 \alpha  (-t_s+\chi )^{-1+\alpha }\dot{\chi}^2}{\kappa ^2}}}
\, .
\end{align}
As in the previous steps, for the choice of the parameters given in Eq. (\ref{parameters}), the parameter $\ddot{\sigma}$ can be approximated by $\ddot{\sigma}\simeq \ddot{\varphi}$. So finally the second slow-roll index $\eta_V$ reads,
\begin{equation}\label{someteimeifeel}
\eta\simeq \epsilon -\frac{\ddot{\varphi}}{H\dot{\varphi}}\, .
\end{equation}
But as was shown in \cite{barrowslowroll}, the second term of Eq. (\ref{someteimeifeel}), is the second Hubble slow-roll index for the canonical scalar field, that is, $\eta_H=-\frac{\ddot{\varphi}}{H\dot{\varphi}}$, and therefore the we get,
\begin{equation}\label{exactprove}
\eta=\epsilon+\eta_H\, .
\end{equation}
As was explained in the previous section and also was shown in \cite{barrowslowroll}, the expression $\epsilon+\eta_H$ is equal to the potential slow-roll parameter $\eta_V$ corresponding to the single scalar field $\varphi$ (see Eq. (\ref{hgfgsdgs})). Therefore we proved that with the choice of the parameters made in Eq. (\ref{parameters}), the two scalar field potential slow-roll index $\eta$ is actually equal to the single scalar field slow-roll parameter $\eta_V$, which we defined in the previous section. This result validates our original claim that the two scalar theory is dynamically equivalent to a single scalar theory. In practise, we chose a trajectory for which the potential and kinetic term of the field $\chi$ could be disregarded, which is equivalent to a single scalar field trajectory. Therefore, at the observational indices level, the single scalar field trajectory and the two scalar theory with the parameters chosen as in Eq. (\ref{parameters}). However potential differences between the two approaches can be revealed because the small contribution of the scalar field $\chi$ can generate non-Gaussianities. This study is quite interesting but exceeds the purposes of this article. Before closing this subsection we need to note that the spectral index of primordial curvature perturbations and the scalar-to-tensor ratio, corresponding to the slow-roll indices we calculated in this section, are equal to,
\begin{equation}\label{hdhdfggjuishuii}
n_s=1-6\epsilon+2\eta,\,\,\, r=16 \epsilon\, ,
\end{equation}
For the Hubble slow-roll parameters this relation is different, as was shown in \cite{barrowslowroll}.

\subsubsection{Hubble Slow-Roll Parameters Approach}

In this section, we adopt a different approach in comparison to the one we adopted in the previous section. Specifically, the main difference will be in the definition of the second slow-roll parameter $\eta$, since the definition of the first slow-roll parameter coincides with the one we used previously. Particularly, we shall use the Hubble slow-roll parameter $\eta_H$, which is equal to \cite{barrowslowroll}:
\begin{equation}\label{gfhsgs}
\eta_H=-\frac{\ddot{H}}{2\dot{H}H}\, .
\end{equation}
We can easily write the Hubble slow-roll parameter $\eta_H$ in terms of the $\sigma$ field we introduced in the previous section, and by using the FRW (\ref{fieldhf}), we get,
\begin{equation}\label{eqnsdfger}
\ddot{H}=-\kappa^2 \dot{\sigma} \ddot{\sigma }\, .
\end{equation}
Then by combining Eqs. (\ref{eqnsdfger}), (\ref{fieldhf}) and (\ref{voithitiki}), the second Hubble slow-roll parameter $\eta_H$ can be written as follows,
\begin{equation}\label{sechubprolept}
\eta_H=-\frac{\ddot{\sigma}}{\dot{\sigma}H}\, .
\end{equation}
By using Eqs. (\ref{dhfhf}) and (\ref{voithia2}), the parameter $\eta_H$ can be written as follows,
\begin{equation}\label{ghybbleeqn}
\eta_H=-\frac{-4 c_2 \alpha  (-t_s+\chi )^{-1+\alpha }\ddot{\chi}\dot{\chi}+2\dot{\varphi}\ddot{\varphi}}{2H\left(\dot{\varphi}^2-\frac{2 c_2 \alpha  (-t_s+\chi )^{-1+\alpha }\dot{\chi}^2}{\kappa ^2}-4 c_2 (-1+\alpha ) \alpha  (-t_s+\chi )^{-2+\alpha } \dot{\chi}\right) }\, ,
\end{equation}
and by using the values of the parameters given in Eq. (\ref{parameters}), the parameter $\eta_H$ can be simplified as follows,
\begin{equation}\label{hdwahedsil}
\eta_H\simeq -\frac{\ddot{\varphi}}{H\dot{\varphi}}\, .
\end{equation}
By looking Eq. (\ref{hdwahedsil}), we realize that this is nothing else but the Hubble slow-roll parameter corresponding to a single canonical scalar field $\varphi$ (see previous section). Therefore in this case too, the two scalar-field formalism ends up to the single scalar field description, if the values of the parameters are chosen as in Eq. (\ref{parameters}). Note that in the present case, the spectral index of primordial curvature perturbations $n_s$ is equal to \cite{barrowslowroll},
\begin{equation}\label{shedpectralperturb}
n_s=1-4\epsilon +2\eta_H \, .
\end{equation}
Also we need to stress that if we substitute the first order in the slow-roll expansion result $n_H=-\epsilon+\eta$, then Eq. (\ref{shedpectralperturb}) becomes identical to Eq. (\ref{hdhdfggjuishuii}).

As we already stated, this result is strictly dependent on the fine-tuned choice of the parameters (\ref{parameters}). Therefore, if we can safely neglect the kinetic term and the contribution of the scalar field $\chi$ to the potential, then our results are valid. However, some differences between the two scalar field theory and the single scalar field theory can be observed when non-Gaussianities are considered. We defer this task to a future publication.

\section{Discussion and conclusions}

In the previous section we mentioned that within the theoretical framework of general scalar-tensor theories, if the FRW equations are combined with the scalar field equation of motion, these can be written in terms of a dynamical system. The solution of the reconstruction method, is a solution to this dynamical system, so the stability of this solution may reveal if the reconstruction solution is stable towards linear perturbations. Here, we shall discuss the possibility of having an instability in the dynamical system and we shall offer another perspective of having an instability in a theory containing a finite time Type IV singularity. In principle, the existence of instabilities may be considered as an unwanted feature, but the existence of the instabilities may possibly indicate the point at which a new physical phenomenon takes place. This is true, since at the instability point, the classical trajectory that the dynamical system followed is unstable, and therefore the system becomes strongly unstable. The important thing about the Type IV singularities is that the dynamical system of the scalar fields becomes unstable at exactly the time at which the singularity occurs. Therefore, if we appropriately choose this time, it is possible to explain indirectly why physical phenomena described by the dynamical system, at the point of the singularity need a new physical description yet to be found. An example on this account could be for example the stopping of inflation in certain scalar theories of inflation. We hope to address this issue soon and work is in progress.

In conclusion, in this paper we constructed a theoretical framework consisting of two scalar fields, in order to support a singular evolution of the Universe with a nearly $R^2$ inflation potential. The singularity we incorporated in the evolution is a Type IV singularity, which we assumed it occurs at the end of the inflationary era. The model we used consisted of two scalar fields, one canonical and one non-canonical, with the canonical scalar being the one describing the nearly $R^2$ inflation potential. Near the singularity, and by appropriately choosing the parameters, the canonical scalar field dominates the evolution of the Universe, since the non-canonical scalar field is chosen in such a way so that it's contribution at early and intermediate times is negligible, in reference to the contribution of the canonical scalar. In addition, we assumed that the non-canonical scalar does not satisfy the slow-roll condition, and therefore it's contribution becomes significant and it dominates the Universe's evolution at late-time. As we showed, in some cases, the scalar model we used has some qualitatively appealing attributes, with regards to the cosmological evolution. Particularly, it is possible that the Type IV singularity that occurs at the end of inflation has a direct impact on the late-time dark energy era. Specifically, the late-time acceleration is driven by the Type IV singularity and moreover the dark energy era is a nearly phantom era, infinitely close however to a de Sitter expansion. 

Having found which nearly $R^2$ inflation potentials can incorporate the Type IV singularity in their theoretical framework, we found the corresponding Jordan frame pure $F(R)$ gravity and also worked out the stability of the cosmological solution of the reconstruction method we used. In addition, we studied certain limiting cases of $R+R^p$ theories in the Einstein frame and investigated how these theories can incorporate a Type IV singularity. 

In fact, we demonstrated that singular inflation is quite possible, while transitions between singular and its companion non-singular inflation is not easy to realize. In the absence of a complete quantum theory of gravity, which simultaneously incorporate the singularities in a consistent, complete and unified theoretical framework, we believe our work serves as another step towards the understanding of the nature of these singularities and more importantly, of their impact on the Universe's current evolution. In view of the fact that what we actually see now is the remnants of the primordial quantum theory of gravity, without the quantum phenomena playing an important role, the classical cosmological finite time singularities are useful tools to explore the quantum phenomena within a classical theoretical framework. More importantly, since Type IV singularities are not-catastrophic singularities, like the initial singularity or the Big Rip finite time singularity, these may serve as a doorway to the complete noesis of the quantum nature of cosmological phenomena. In fact, in Loop Quantum Cosmology \cite{LQC}, the initial singularity does not occur, but finite time singularities may occur. The Type IV singularities do not cause geodesics incompleteness of spacetime, so in principle they are harmless, but however the complete nature of these has to be understood, since these may have direct impact on observational indices \cite{Nojiri:2015fra}. It is possible that these singularities may be responsible for ending acceleration or deceleration eras, through instabilities they cause. The indicators of this effect may be the instabilities that often occur in the theories described by one scalar or even two scalar fields. So practically an instability that occurs at the time when the singularity occurs may practically signal the change of the evolutionary mechanism at that point, see for example Ref. \cite{oikonomouodigrexit}.  

Finally, a last remark for the model we used in this paper. Basically, it is the $R^2$ inflation model in the Einstein frame plus an arbitrary but the simplest deformation, that makes the evolution singular. The models we studied cannot be considered as toy models, since these are realistic models, with regards to the canonical scalar field $\varphi$, and the $\chi$ part can be chosen in a simple way so that singular evolution is realized. So practically the $\varphi$ was chosen on the basis that the $R^2$ inflation model is in agreement with current observational data \cite{planck}. As we demonstrated, at the level of observational indices, the scalar field $\chi$ does not contribute significantly at early times, while at late-times it dominates the evolution. A quite interesting task to investigate however, is to investigate the effects of the second scalar fields on the non-Gaussianities of the model. Particularly it is possible that the second scalar could generate non-Gaussianities at early-time, although it makes minor contribution to the early-time evolution. We hope to address this issue soon.

\section*{Acknowledgments}

The work of (S.D.O.) is supported by MINECO (Spain), projects FIS2010-15640 
and FIS2013-44881 and by Russian Ministry of Education and Science (S.D.O).

\end{document}